\documentclass[amsmath,nofootinbib,twocolumn,superscriptaddress]{revtex4-1}
\pdfoutput=1
\usepackage[colorlinks=true,citecolor=blue,urlcolor=blue]{hyperref}
\usepackage{aas_macros,graphicx,marvosym,subfigure,amssymb,amsmath}
\usepackage{comment}
\usepackage{slashed}

\newcommand{\pd}{\,\partial}
\newcommand{\orderof}[1]{O(#1)}
\newcommand{\abs}[1]{\lvert #1 \rvert}

\DeclareMathAlphabet{\mathcalligra}{T1}{calligra}{m}{n}

\usepackage{comment}
\DeclareMathAlphabet{\mathpzc}{OT1}{pzc}{m}{it}
\def\qp{\ensuremath{\mathcalligra{q}}}

\newcommand{\Lie}{\mathcal{L}}

\newcommand{\bs}[1]{\boldsymbol{#1}}

\def\brs{\ensuremath{s}}
\def\brx{x}
\def\brt{t}

\begin{document}
\title{Horizon instability of extremal Reissner-Nordstr\"om black holes to charged perturbations}
\author{Peter~Zimmerman\footnote{{\tt peterzimmerman@email.arizona.edu}}}
\affiliation{Department of Physics, University of Arizona}
\date{\today}
\begin{abstract}

We investigate the stability of highly charged Reissner-Nordstr\"om black holes to charged scalar perturbations.   We show that the near-horizon region exhibits a transient instability which becomes the Aretakis instability in the extremal limit. The rates we obtain match the enhanced rates for nonaxisymmetric perturbations of the near-extremal and extremal  Kerr solutions.  The agreement is shown to arise from a shared near-horizon symmetry of the two scenarios.

\end{abstract}
\maketitle

\section{Introduction}
The importance of black holes in modern theoretical physics cannot be denied. At large scales, numerous astrophysical phenomena feature black holes as major players: stellar mass black holes are the end states of ``high-mass'' stellar collapse \cite{OppenheimerSnyder:1939}, supermassive black holes have the capacity to source high-energy jets seen in the centers of galaxies \cite{BlandfordZnajek:1977}, and binary black holes provide a  source of  gravitational waves, recently detected for the first time \cite{TheLIGOScientific:2016qqj}.  At small scales, black holes provide a testing ground for theories of quantum gravity, which must reproduce the area law of entropy \cite{Bekenstein:1973ur,Hawking:1974sw} at the semiclassical level.  
 Mathematically, the nonlinear stability of black holes is far from understood, though the linear stability problem of black hole solutions has seen tremendous progress in recent years \cite{Dafermos:2016uzj,Dafermos:2014cua,Finster:2016tky}. 

Solutions describing black holes near their extremal limits have attracted additional interest. This may be due in part to the fact that in the extremal limit many problems become analytically tractable.  For instance, microstate counting simplifies near the BPS (extremal) bound where certain fields behave classically. Hence, extremal solutions have been important for string theory calculations of black hole entropy \cite{Strominger1996}.  Apart from their analytical tractability \cite{TeukolskyPress1974,porfyriadis-strominger2014,hadar-porfyriadis-strominger2014,lupsasca-rodriguez-strominger2014,zhang-yang-lehner2014,lupsasca-rodriguez2015,hadar-porfyriadis-strominger2015,GPW15,gralla-lupsasca-strominger2016,compere-oliveri2016,porfyriadis-shi-strominger2016}, extremal black holes have uniquely interesting physical features. These include turbulentlike dynamics \cite{Yang:2014tla}, unique telltale observational features \cite{CunninghamBardeen1972,Andersson2000,Glampedakis2001,Yang:2013uba,Gralla:2016qfw,Burko:2016sfi}, and enhanced symmetries shared by certain conformal field theories  \cite{kerrCFT}.  

A new development came in 2010 when Aretakis proved that extremal horizons are linearly unstable \cite{Aretakis:2010gd,Aretakis:2011ha,Aretakis:2011hc}.  For perturbations of nonextremal black holes the redshift of outgoing radiation at the horizon controls transverse derivatives occurring in energy estimates \cite{Dafermos:2005eh}.  However,
the redshift factor degenerates in the extremal limit.  The lack of a redshift, combined with the existence of conserved quantities on the future horizon \cite{Aretakis:2013oja},  generically gives rise unbounded polynomial growth of sufficiently high-order derivatives at late times \cite{Lucietti2012,Aretakis:2012ei,Murata:2012ct}.  
The polynomial growth was shown to arise in a mode expansion from a branch point in the Laplace transform at the superradiant frequency \cite{Casals:2016mel}.

Physically, extremal solutions occupy a set of measure zero and can never be realized thermodymically in finite time \cite{Israel:1986gqz}. However, given the continuous dependence of the Kerr solution on the spin parameter, 
  the Aretakis ``instability'' is exhibited in a tamer form for near-extremal solutions as well \cite{Murata:2013daa,Gralla:2016sxp}. 

Here we consider charged (denoted $q$) massless scalar field perturbations $\psi$ 
of a Reissner-Nordstr\"om (RN) black hole,  the unique static electrovacuum solution in four dimensions.

There are many reasons to study charged perturbations of the RN spacetime.
For one,  the highly charged RN scenario provides a simplified setting which captures many features of the  rapidly spinning astrophysical Kerr black hole.   
In this paper we show that the linearized near-horizon dynamics of RN, including the horizon instability, 
share many salient features with Kerr and can be understood within background spherical symmetry.   
These shared features include a set of long-lived modes and an Aretakis rate enhancement relative to neutral scalar perturbations that is identical to the nonaxisymmetric rate enhancement found in \cite{Casals:2016mel} for perturbations of Kerr. In fact, many results carry over \emph{exactly} to the Kerr analysis by simply mapping $\qp \to m$, where $m$ is the azimuthal mode number of a Kerr perturbation. Moreover, the (near-)extremal RN geometry offers an ideal setup to study nonlinear interactions between long-lived modes. Such a study may unearth more evidence of gravitational turbulence in asymptotically flat spaces \cite{Yang:2014tla},  pushing the limits of the established fluid gravity correspondence \cite{Baier:2007ix, Bhattacharyya:2008jc} wherein anti–de Sitter (AdS) black branes have been shown to exhibit turbulent features \cite{Adams:2013vsa, Green:2013zba}. 
Yet another reason to study charged fields in (near)-extremal Reissner-Nordstr\"om, when allowing AdS asymptotics,  comes from the AdS/CMT correspondence \cite{Hartnoll:2009sz}, as it may inform our understanding of novel low-temperature strongly coupled states of matter \cite{Gubser:2008px,Hartnoll:2008vx}. 

The scalar field $\psi$ is decomposed into a set of discrete spherical harmonic modes $(\ell ,m )$ on $\mathbb S^2$. It is also meromorphically extended into the complex frequency plane through a Laplace transform, which we invert to resolve the time dependence.  The near-horizon field modes are shown to organize into spaces (modules) imaged by the action
of  the  near-horizon symmetry group. 
The representations are labeled in part by the so-called ``conformal weight'' $h$.
 Through the field equations, we find that the conformal weight is related to the  (eigenvalues of the) $SO(3)$ and $U(1)$ Casimirs, $K_\ell := \ell(\ell+1)$ and $\qp^2$, by
\begin{equation}\label{eq: h}
 h = \frac12 + \sqrt{ \frac14 + K_\ell  - \qp^2 }.
\end{equation}  
Our main results apply to modes for which $h \in \mathbb C$.  In the language of representation theory, these are the ``principal series'' representations \cite{Barut:1965}.  The principal representations were shown to dominate the near-horizon response of the rapidly spinning Kerr black hole \cite{Gralla:2016sxp}, where the azimuthal mode number $m$ plays the role of $\qp$ in \eqref{eq: h}.  In the charged case studied here, we again find that the principal modes run the show, giving rise to horizon dynamics replicating extremal Kerr. 

One explanation of the similarities with the Kerr scenario lies in the shared presence of a critical frequency at which the horizon flux is zero---the superradiant bound.
The perturbation spectra of both (near-extremal) spacetimes 
feature a 
family of weakly damped quasinormal modes clustered around the superradiant bound frequency \cite{Detweiler1977,Leaver1985, Glampedakis2001, Cardoso2004, Hod2010chargedfields, Hod2012349, HodEikonal2012,Yang:2013uba, Cook:2014cta, Richartz:2014jla, PhysRevD.88.024054, Richartz:2016}.  
The collective excitation of these slowly decaying quasinormal modes gives rise to a steeply graded field with large local energy density  measured near the horizon.
 In the extremal limit, a confluence of these resonant modes forms a branch point precisely at the superradiant bound frequency.  In this paper we analytically compute the field arising from 
the long-lived quasinormal modes (near-extremal case)  and branch cut integral (extremal case), presenting decay rates for each.

We follow the conventions of \cite{MTW} and use geometric units $G = c = 1$.

\section{Field equations}
We consider the charged scalar field
\begin{equation}\label{eq:scalar wave}
    D_\alpha D^\alpha \psi  = 0,
\end{equation}
where $D_\alpha = \nabla_\alpha - i q A_\alpha$, as a linear perturbation of  the Reissner-Nordstr\"om spacetime. The field $\psi$ is constructed via the Green function governing the linear response to suitable initial data or a prescribed compact source.  The Green function is defined as the distributional solution of the adjoint equation
\begin{equation}\label{eq:Green eqn}
 {D_{\alpha} }^* {D^{\alpha}}{}^* G(X,X') = \delta_4(X,X'),
\end{equation}
where $\delta_4(X,X')$ is the covariant Dirac distribution and  ``$*$'' signifies taking the complex conjugate. We choose $G(X,X')$ to vanish when the spacetime point $X$ is not in the chronological future of $X'$.
With this choice, the Kirchhoff representation of the ``forward'' solution in the absence of sources is
\begin{equation}\label{eq:Kirchhoff}
 \psi = \int_{\Sigma} \left (\psi_\Sigma D^*_\alpha G^* -  G^* \, D_{\alpha}\, \psi_{\Sigma} \right) n^\alpha \sqrt{h}\,  d^3 y
\end{equation}
for given compact support initial data $\psi_\Sigma$ and $ n^\alpha D_\alpha \psi_\Sigma$ on the spacelike surface $\Sigma$ with future directed normal $n^\alpha$ and intrinsic metric $h_{jk}$. We restrict to  initial data supported away from the future outer horizon.

 In ingoing coordinates $x^\mu = (v,r,\theta,\phi)$, the background metric and gauge field are given by
\begin{align}
    ds^2 &= -f(r) dv^2 + 2 dv dr + r^2 d \Omega^2 \label{eq:background met} \\
    A &= - \frac{Q}{r} dv,\label{eq:background A}
\end{align}
where $f(r) = 1 -2M/r +Q^2/r^2 = (r-r_-)(r-r_+)/r^2$ with $r_+$ and $r_-$ denoting the locations of the inner and outer horizons, $r_{\pm} = M \pm \sqrt{M^2 - Q^2}$. The quantities $Q$ and $M$ are the ADM charge and mass of the black hole, respectively.

To separate variables \footnote{Here we are separating the adjoint of \eqref{eq:Green eqn}.}
 we adopt a mode decomposition
\begin{align}
G^* = \tfrac{1}{2\pi} &\sum_{\ell m} Y_{\ell m}(\theta, \phi) Y^{*}_{\ell m}(\theta',\phi')\nonumber \\ & \times \int_{ i c- \infty }^{ic+\infty}  e^{- i \omega v}  
 \tilde g(r,r') \mu(r') d \omega,
\end{align}
where $Y_{\ell m}$ are spherical harmonics satisfying $\Delta_{\mathbb {S}^2} Y_{\ell m} = -K_\ell Y_{\ell m}$ and  $\tilde g(r,r')$ is the ``transfer function''  obeying  
\begin{align}
 f & r^2  \pd_r^2 \tilde g(r,r')  + 2\left(r-M -ir(\omega r - q Q) \right)\pd_r \tilde g(r,r') \nonumber \\ &+ \left(-K_\ell+ i (q Q - 2 \omega r )\right) \tilde g(r,r')= \mu^{-1}(r) \delta(r-r').
\end{align}
We have also introduced a ``weight factor'' given by 
\begin{align}
\mu(r) &= \frac{1}{r^2 f} \exp \left( \int 2\left(r-M -ir(\omega r - q Q) \right)/(r^2 f)\,dr \right) \\
&= e^{-2 i \omega r_*} \exp \left (2i   qQ \int \frac{dr}{rf} \right) \\
&\sim e^{-2 i \omega r_*} (r/r_+-1)^{2i qQ}, \,\quad \, r \to \infty,
\end{align}
where $r_* = \int dr/f$.\footnote{The function $\mu$ is a weight in the following sense. Define $E := f r^2  \pd_r^2  + 2\left(r-M -ir(\omega r - q Q) \right)\pd_r  \nonumber + \left(-K_\ell+ i (q Q - 2 \omega r )\right)$. Then $E$ is symmetric in the bilinear $(a,b) = \int a(r) b(r) \mu(r)dr$ in the sense that $(a,Eb)=(Ea,b)$.} The quantity $c$  is a small positive constant chosen to put the integration contour in a strip where $\tilde g$ is holomorphic.

A suitable transfer function consistent with the causal conditions imposed on $G$ may be constructed from homogeneous solutions $R_{\rm in}$ and $R_{\rm up}$, which are defined as having no incoming radiation  from the past horizon (in) and past null infinity (up).  To construct this quantity explicitly, we employ the ``variation of parameters'' formula 
\begin{equation}\label{eq:transferDef}
 \tilde g (r,r')  = \frac{R_{\rm in}(r_<) R_{\rm up}(r_>) }{ \mathcal W},
\end{equation}
making use of the notation $r_< := \min(r,r')$, $r_> := \max(r,r')$, and introducing the scaled Wronskian 
 $\mathcal W  = r^2 f\mu(r) \left(R^{\rm in} \pd_r R^{\rm up}  - R^{\rm up}  \pd_r R^{\rm in}\right)$  such that  $\pd_r \mathcal W=0$.  
 As defined, the in solution is regular on the future horizon and the up solution is regular at  infinity. 
 
Following \cite{TeukolskyPress1974},  homogeneous radial solutions are obtained in two asymptotic regions, the ``near'' and ``far'' zones, and then matched in a region of common overlap: the ``buffer.''  The existence of a buffer region relies on the presence of an additional parameter deemed small. When the black hole is extremal, a sufficiently parameter arises in our analysis as the deviation of the frequency of the perturbation from the superradiant bound frequency. This is a simple but profound consequence of the fact that waves of all frequencies defined with respect to the near-horizon geometry limit to the superradiant bound when viewed in coordinates adapted to the asymptotically flat region. We thus pattern our analysis on \cite{porfyriadis-strominger2014}, obtaining asymptotic solutions by solving the radial equation directly in the limiting spacetimes (limiting sections on the $U(1)$ bundle).  In doing so we make use of the near-horizon symmetries to streamline our derivation of the Aretakis instability. 

 \section{Homogeneous solutions}
 
 \subsection{Near-zone solutions}
 \subsubsection{Near-extremal near-zone solution}
We now derive the ingoing near-zone solution by solving the radial equation near the horizon and imposing no incoming radiation.  
 For near-extremal solutions
\begin{align}\label{eq:sigma}
    \sigma := \frac{ r_+ - r_-}{r_+} \ll 1.
\end{align}
We further introduce  a shifted radial coordinate 
\begin{equation}
    x := \frac{r-r_+}{r_+},
\end{equation}
which puts the outer horizon at zero, forming the inner boundary of our working domain.

To begin, we construct the near-horizon geometry by introducing a scaling parameter $\lambda$. For the near-extremal RN black hole, the near-horizon geometry of interest is the end point $(\lambda \to 0)$ of a flow along a one-parameter family of spacetimes having $ \sigma = \lambda \bar \sigma $. The flow is taken at fixed ``scaled'' coordinates $\bar v = \lambda v / r_+$, $\bar x = x/ \lambda$, and $\bar \sigma$. The limiting metric forms a patch of the Robertson-Bertotti universe, $ \mathrm{AdS}_2 \times \mathbb{S}^2$, with metric \cite{robinson1959solution,Bertotti1959, Maldacena:1998uz, Carter2009}
\begin{equation}\label{eq: near-nhern}
 r_+^{-2} ds^2 = - \bar x (\bar x + \bar \sigma) d\bar v^2 + 2 d \bar v d \bar x + d \Omega^2.
\end{equation}
The gauge field in the form \eqref{eq:background A} is singular in this limit. 
We choose a section on the $U(1)$ bundle compatible with the near-horizon limit by shifting $A \to A + d \chi$, where $\chi = \frac{Q}{r_+}v, $  such that $A(r_+)=0$. The limit $\lambda \to 0$ gives 
\begin{equation}\label{eq: A nearNHERN}
\bar A = Q \bar x  \, d \bar v.
\end{equation} 
The near-zone scalar field frequency in this gauge is then given by 
\begin{equation}\label{eq:NHF}
    \bar \omega  := \frac{\bar \sigma r_+(\omega - q  \Phi_+) }{ \sigma} , \quad \Phi_+ := \frac{Q}{r_+}.
\end{equation}

Upon separating variables using $e^{- i \bar \omega \bar v} R(\bar x) Y_{\ell m}(\theta,\phi)$, the homogeneous radial equation is found to be  an ordinary hypergeometric equation
\begin{equation}\label{eq:near ode}
 \bar x \left( \bar x + \bar \sigma\right) R'' + \left( \bar \sigma -2 i (\bar \omega + \qp \bar x + i \bar x )\right)R' - \left( K_\ell + i \qp \right) R = 0,
\end{equation}
where $\qp := r_+ q$.
Of the two linearly independent solutions we choose the ingoing one 
\begin{equation}
 R_{\rm near}^{\rm in} = {}_2 F_1 \left(1-h-i \qp,h-i \qp,1-i\bar k,-\frac{\bar x}{\bar \sigma} \right),
\end{equation}
where we have introduced 
\[ \bar k:= 2 r_+(\omega - q  \Phi_+) / \sigma. \]
Using  Eq.~(15.8.2)  of \cite{nist},
we find that the ingoing solution has the large-$\bar x$ (buffer-zone) asymptotics
\begin{equation}\label{eq:Rin buffer ingoing}
 R_{\rm near}^{\rm in } \sim
\bar A (\bar x/\bar \sigma)^{h-1+ i \qp}+ \bar B ( \bar x/ \bar \sigma)^{-h+ i \qp}, \quad \bar x \to \infty,
\end{equation}
where 
\begin{subequations}
\begin{align}\label{eq: ne buffer coeffs}
 \bar A &= \frac{\Gamma(2h-1)\Gamma(1-i\bar k)}{\Gamma(h- i \qp) \Gamma(h-i\bar k+ i \qp) }, \\
 \bar B &= \bar A(h\leftrightarrow 1-h). \label{eq:bar B}
\end{align}
\end{subequations}
The operational meaning of $h \leftrightarrow 1-h$ in \eqref{eq:bar B}  is to interchange $h$ with $1-h$ everywhere in $A$. 

\subsubsection{Extremal near-zone solution}\label{sec:extremal limit}
We are also interested in seeing what happens in the extremal limit.
To study this regime, we require  the near-horizon radial functions. These functions are solutions to the radial differential equation derived from the field equation in the extremal near-horizon geometry.  
 The extremal near-horizon limit is obtained by the same limiting procedure used in the previous section, but now we instead fix the scaled coordinates 
\begin{equation}
 \hat x = \frac{x}{\lambda^p}, \quad \hat v = \frac{\lambda^p v}{r_+}
\end{equation}
with $p \in (0,1)$ and take $\lambda \to 0$ while keeping $\bar \sigma $ fixed. The extremal limiting geometry forms a different patch of  $\mathrm{AdS}_2 \times \mathbb{S}^2$ with metric 
\begin{equation}\label{eq:NHERN}
  r_+^{-2}ds^2 = - \hat x^2 d\hat v^2 +  2 d \hat v d \hat x + d \Omega^2,
\end{equation}
which is diffeomorphic to the near-extremal patch \eqref{eq: near-nhern} \cite{Maldacena:1998uz,Spradlin:1999bn}.

 In the frequency domain, one also must hold
  \begin{equation*} \hat \omega = \lambda^{-p} k,\quad \mathrm{ where} \quad k := r_+ (\omega - q \Phi_+), \end{equation*}  
fixed  to  have a well-defined Laplace transform. Then, the mode functions $e^{-i \hat\omega  \hat x} R(\hat x) Y_{\ell m}(\theta,\phi)$ reduce the extremal near-horizon field equation to
\begin{equation}\label{eq:ext near-ode}
 \pd_{\hat x}(\hat x^2 \pd_{\hat x} R ) -2 i(\hat \omega + \qp \hat x) \pd_{\hat x} R - (K_{\ell} +i \qp)R =0.
\end{equation} 
Notice that the regular singular points at the inner and outer horizon which determined the solution of the near-extremal near-horizon radial equation \eqref{eq:near ode} have ``come together'' to form an irregular singular point at the degenerate horizon $\hat x =0$. The confluence of singular points may be exploited to generate 
the ingoing solution of \eqref{eq:ext near-ode} from the previously obtained ingoing (near-extremal) near-zone solution by taking the scaling limit
\begin{equation}
\lim_{\lambda \to 0} {}_2 F_1(1-h- i \qp, h- i \qp, 1- 2 i \lambda^{p-1} \hat \omega/\bar \sigma, - \lambda^{p-1} \hat x/\bar\sigma)   
\end{equation}
and using the confluence identity \cite{GPW15} $W_{\nu,\mu}(z) = \lim_{c\to\infty} {}_2 F_{1}\left(\mu-\nu-1/2,1/2-\mu-\nu,c;1-c/z\right) e^{-z/2} z^{\nu}$, where $W_{\nu,\mu}(z)$ is the Whittaker confluent hypergeometric function. It is easily checked that the result
\begin{equation}\label{eq:Rin extremal}
 %
 R_{\rm near}^{\rm in} (\hat x) = \left(- \frac{2 i \hat \omega }{\hat x} \right)^{- i \qp} \exp{\left(- \frac{i \hat \omega}{\hat x } \right)} W_{i \qp, h-1/2}\left( - \frac{2 i \hat \omega}{\hat x} \right)
\end{equation}
satisfies \eqref{eq:ext near-ode}.
 In the neighborhood of the horizon, the asymptotic ingoing solution is given by 
\cite{nist}
\begin{equation}\label{eq:ingoing extreme hor} 
 R_{\rm near}^{\rm in} \sim \sum_{j=0}^\infty \frac{ \left(h- i \qp \right)_j \left(1-h - i \qp \right)_j }{ j ! }
 \left( \frac{  \hat x }{ 2 i \hat \omega } \right)^j, \quad \hat x \to 0,
\end{equation}
where $(a)_j := \Gamma(a+j)/\Gamma(a)$ is the Pochhammer symbol. To find the buffer-zone asymptotics, one may either take the scaling limit $\lambda \to 0$ of Eq.~\eqref{eq:Rin buffer ingoing} at fixed hatted coordinates and with fixed finite $\hat \omega$, or equivalently, employ the large-$\hat x$ asymptotics of \eqref{eq:ingoing extreme hor} directly via Eq.~(13.14.18) of \cite{nist}.  Taking the former route, and applying Eq.~(5.11.12) of \cite{nist}, one finds
\begin{equation}\label{eq:ingoing extreme buff}
 R_{\rm near}^{\rm in}  \sim \hat A \, \hat x^{h-1+ i \qp} + \hat B\, \hat x^{-h + i \qp}, \qquad \hat x \to \infty,  
\end{equation}
where 
\begin{subequations}
\begin{align}\label{eq:extreme buffer coeffs}
 \hat A &= \frac{ \Gamma(2h-1) }{\Gamma(h- i \qp ) } \left( - 2 i \hat \omega \right)^{1-h-i \qp} , \\
 \hat B &= \hat A(h \leftrightarrow 1-h).
\end{align}
\end{subequations}
Notice that the extremal buffer-zone coefficients exhibit branch points at the superradiant bound $\hat \omega=0$.

\subsection{Far-zone solutions}
To obtain the far-horizon limit, we fix the far (asymptotic) coordinates $x^\mu=(v,r,\theta,\phi)$ and take $ \sigma \to 0$ and $k=r_+(\omega - q \Phi_+) \to 0$.  This limit corresponds to  perturbations of the extremal geometry at the critical frequency $k=0$.

In this limit, the inner and outer horizons (singular points of the ODE) ``converge" and the radial equation is
\begin{equation}\label{eq:far-eq ingoing}
    (x^2 R')' - 2 i \qp x (1+x) R' - \left(K_\ell + i \qp(1+2x) \right) R = 0.
\end{equation}
 Equation ~\eqref{eq:far-eq ingoing} is a confluent hypergeometric differential equation with solution 
\begin{align}\label{eq:far}
    R_{\rm far} = & P x^{h-1+ i \qp} {}_1 F_1(h+ i \qp,2h, 2 i \qp x) \nonumber \\ & +  x^{-h + i \qp} \, {}_1 F_1(1-h+i\qp,2-2h,2i\qp x).
\end{align}
Again, we have used that $h(h-1) = K_\ell - \qp^2 $. The far solution is regular for the choice  
\begin{equation}\label{eq:PtoQ ingoing}
P = -\frac{ (-2i \qp)^{2 h-1} \Gamma (2-2 h) \Gamma (h-i \qp)}{\Gamma (2 h) \Gamma (1-h-i \qp)}.
\end{equation}
The buffer-zone asymptotics are obtained using ${}_1 F_1 (a,b,0) = 1$ for any $a$ and $b \notin \mathbb{Z}^{-} \cup \{0\}$. 
One may verify that the large-$\bar x$ asymptotic near-extremal solution, as given in Eq.~\eqref{eq:Rin buffer ingoing}, matches  onto the far-horizon solutions just obtained. Moreover, the far-zone solution matches onto the hatted solution Eq.~\eqref{eq:ingoing extreme buff},
as can be seen by reintroducing $k := \lambda^p \hat \omega$ in the near-horizon solution and replacing $\hat A \hat x^{h-1+ i \qp}$ by $A  (x/k)^{h-1+ i \qp}$ (likewise for the $\hat B$ term).

\section{Linear response: charged field instability}
The time dependence of the Green function has an intricate relationship with the analytic structure of the transfer function $\tilde g$ and the contour of the inversion integral \eqref{eq:Green eqn} \cite{Leaver1986b}.  Subtleties aside, the qualitative picture goes as follows. At ``early'' times, direct propagation on the future light cone derives from the large-$\abs{\omega}$ arc.  At very late times, the field exhibits a power-law time dependence deriving from the branch point(s) located at $\omega=0$  for the nonextremal black hole,  and $\omega= q \Phi_+$ and $\omega = 0$ at extremality. At ``intermediate'' times, the field takes the form of  
a decaying sinusoid coming from the poles of the transfer function, the quasinormal modes.
\vspace{-.1cm}
\subsection{Near extremal case}

\subsubsection{Quasinormal mode spectrum}
The spectrum of long-lived quasinormal modes which characterize the dominant near-horizon ringing is now obtained. 
 From Eq.~\eqref{eq:transferDef} we see that for holomorphic homogeneous solutions, the poles of the transfer function are given by the zeros of the Wronskian $\mathcal W$ alone.
Matching the near and far solutions and demanding linear dependence ($\mathcal{W}=0$) gives the quasinormal mode condition $\bar A  =  P\bar B$, or
{\small
\begin{equation}\label{eq:QNM condition}
 \frac{\Gamma^2(1-2h)\Gamma^2(h-i\qp)\Gamma(h+i\qp-i\bar k)}{\Gamma^2(2h-1)\Gamma^2(1-h-i\qp)\Gamma(1-h+i\qp-i\bar k)} = (-2 i  \sigma \qp)^{1-2h}.
\end{equation}}

To find the quasinormal modes, we adopt an ansatz originating with Hod \cite{Hod2009}, which approximates the zeros of $\mathcal W$ by the negative integer poles of certain gamma functions containing $\bar k$ in \eqref{eq:QNM condition}. The deviation of the quasinormal mode pole position from the pole of the gamma function is parametrized by a complex number $\eta$, which weakly depends on the integer overtone index $n$.  When $h \in \mathbb{R}$ the approximation is validated by the perturbative smallness of $\sigma$ and $\eta = \orderof{\sigma^h}$, while  for complex $h$ the Hod approximation has been found to be numerically accurate to within $\abs{\eta} \approx 10^{-3}$ in typical cases \cite{Yang:2012pj,Cook:2014cta}. We find that the near-horizon quasinormal modes fall into two categories
\begin{subequations}\label{eq:NHQNMs}
\begin{align}
 \bar k_n &= \qp + i \left(h-1 -n \right) + \eta, \qquad ({\rm principal}), \label{eq:hod complex h} \\
 \bar k_n &= \qp  + i \left(-h - n \right),\qquad ({\rm supplementary}) ,
\end{align}
\end{subequations}
where $\bar k_n := 2 r_+(\omega_n - q \Phi_+)/ \sigma$ at the frequency $\omega_n$.  The terminology principal and   supplementary has its roots in the classification of $SO(2,1)$ representations to be discussed in some detail in Sec.~\ref{sec:symmetry}. 

In addition to the weakly decaying near-horizon modes, there are also damped ``far-horizon'' quasinormal modes distributed in $\mathbb C$  
away from the superradiant frequency \cite{Richartz:2014jla}.  As these modes have non-$\sigma$-suppressed exponential decay, they have a negligible contribution relative to the near-horizon modes after an inverse-$\sigma$ time scale. 

\subsubsection{Overtone sum}
For simplicity, let us assume that the perturbation develops from compactly supported initial data  with amplitude peaked  deep within  the far zone where $x' \gg 1$. In this case, we approximate $R_{\rm up}(x')$ by its asymptotics 
\begin{equation}
R_{\rm up} \sim C_{\infty} e^{2i \qp x'}(x')^{2 i \qp -1} \quad x' \to \infty.
\end{equation}
At large $x'$ the weight is given by 
\begin{equation}
\mu \sim e^{-2 i \qp x'} (x')^{-2 i \qp}, \quad x' \to \infty
\end{equation}
such that complete $x'$ dependence is asymptotically given by $\mu(x') R_{\rm up}(x') \sim C_{\infty}/x' \, {\rm as} \,\, x' \to \infty$.
 The transmission amplitude $C_{\infty}$ is straightforwardly computed from the asymptotics of \eqref{eq:far} subject to the outgoing condition \eqref{eq:PtoQ ingoing}. 

As in \cite{Gralla:2016sxp}, we define a near-horizon-mode (NHM) Green function due to poles near the superradiant bound, 
\begin{align}\label{eq:GNMs}
G_{\rm NHM} &= - \frac{ i \sigma }{2 r_+ x'}  
 \sum_{\ell m}  C_\infty Y_{\ell m }(\theta, \phi)Y_{\ell m}^*(\theta',\phi') \\ 
&\quad\times
 \sum_{n=0}^{\infty} e^{- i  \bar k_n \bar V} \frac{  \, R_{\rm in}(\bar X; \bar k_n)  }{ d \mathcal W/d \bar k \big\vert_{\bar k=\bar k_n} },\nonumber
\end{align}
where 
\begin{equation}
\bar V :=  \sigma( v-v')/(2r_+) , \qquad \bar X := x /\sigma.
\end{equation}
The values of $\bar k_n := 2 \bar \omega_n / \bar \sigma $ are given by  \eqref{eq:NHQNMs}.

The residues at the QNM frequencies are 
derived from the zeros of the scaled Wronskian 
\begin{align}
\mathcal W & :=   \Gamma(1 -  i \bar k)\left(  \frac{a  \, \sigma^{1-h-i \qp} }{\Gamma(h-i \bar k +  i \qp ) } +  
\frac{ b \, \sigma^{h-i \qp} }{\Gamma(1-h-i \bar k +  i \qp ) } \right), \nonumber
\end{align}
where 
$a = - \Gamma(2h)/\Gamma(h- i \qp)$, $b=\Gamma(2h-2)/\Gamma(1-h-i \qp) \times P $, with $P$ given in Eq.~\eqref{eq:PtoQ ingoing}.
To find the approximate zeros of $\mathcal W$ we substitute the Hod ansatz \eqref{eq:NHQNMs}, differentiate with respect to $\eta$, and drop $\orderof{\eta}$ error terms.  For the dominant principal modes, we find
\begin{equation}\label{eq:dWdk}
\frac{d \mathcal W}{d \bar k} \Big\vert_{\bar k = \bar k_n} 
= b  \, \sigma^{h-i \qp} \, \left( i (-n)^{n+1} n! \right) \Gamma(h-i\qp -n),\,\,\, h \in \mathbb C, 
\end{equation}
where we have used 
$1/\Gamma(-n- i \eta) = - i \eta (-1)^n n! + O(\eta^2)$.
The supplementary modes are obtained from \eqref{eq:dWdk} by swapping $b \leftrightarrow a$ and $h \leftrightarrow 1-h$.

To compute the overtone sum we again follow \cite{Gralla:2016sxp}.  
However, here we restrict to the principal series modes for which 
\begin{align}\label{eq:GNHM2}
G_{\rm NHM} = & \frac{-i}{2 r_+ x'} \sum_{\ell m} \sigma^{1-h+i \qp} \, C_{\infty} b^{-1} \cdot Y_{\ell m}(\theta, \phi) Y^*_{\ell m}(\theta',\phi') \nonumber \\ &\times  \, e^{-i \left(\qp+i(h-1) \right) \bar V} \, \mathcal{S}_P.
\end{align}
The overtone sum
\begin{align}
\mathcal S_P &:= \sum_{n=0}^{\infty} \frac{(-1)^n e^{-n \bar V}}{n! \Gamma(h-i\qp -n)}\nonumber \\ &\quad \times {}_2F_1(1-h-i\qp,h-i\qp,h-i\qp -n, - \bar X)
\end{align}
is  evaluated by invoking the series representation of the hypergeometric function and interchanging the two sums. 
The result of both summations is
\begin{align}
\mathcal S_P &= \frac{ (1-e^{- \bar V})^{h-i\qp-1} }{\Gamma(h-i\qp)} \Big[ 1 + \bar X \left(1-e^{-\bar V} \right) \Big]^{h-1+ i \qp}.
\end{align}
At late times $\bar V \to \infty$, $\mathcal{S}_P \to 1/\Gamma(h-i \qp)$,  the response decays exponentially according to \eqref{eq:GNHM2}. At semiearly times---after a light-crossing time but before the exponential decay ``takes over''---the overtone sum takes the simple form
\begin{equation}\label{eq:Searly}
\mathcal S_P \approx \frac{\bar V^{h- i \qp-1} } {\Gamma(h-i\qp)}\left(1+ \bar X \bar V\right)^{h-1 + i \qp}, \qquad \bar V \ll 1,
\end{equation}
where $\approx$ signifies we have expanded $e^{-\bar V}$ and kept first order in $\bar V$ only. We further overload the $\approx$ symbol by restricting to values of $V$ large enough that the contribution from the large-$\abs\omega$ arc may be ignored, ensuring that the quasinormal mode sum is convergent.
Finally, taking $n$ radial derivatives of \eqref{eq:Searly} one finds 
\begin{equation}\label{eq:dS}
\mathcal{S}_{P}^{(n)} \approx \frac{\bar V^{h-1-i\qp +n}}{\Gamma(h-i\qp)} \left(h-1+i \qp \right)_n \left(1+\bar X \bar V\right)^{h-1+ i \qp -n}.
\end{equation}
Therefore we have demonstrated that $n$ derivatives of a principal harmonic 
 grow at a rate $\bar V^{-1/2+n}$ for $\bar V \ll 1$. The same transitory growth, including the rate, was found for nonaxisymmetric perturbations of the near-extremal rotating Kerr black hole \cite{Gralla:2016sxp}.

\subsection{Extremal case}
The analysis of the previous section suggests unbounded growth   of radial derivatives
on the horizon as $\sigma \to 0$. 
We now show how the Aretakis instability arises at the extremal  horizon as a steep-gradient tail by examining the Laplace transform near the branch point at the superrandiant bound frequency. 

Near the horizon, the asymptotic transfer function for far-zone initial data $(x' \gg 1)$ is given by
\begin{equation}\label{eq:transfer}
\tilde g \sim \frac{R_{\rm up}(x') \, (- 2 i k)^{i \qp} }{a (-2 i k)^{1-h} + b (-2 i k)^{h} }\sum_{j=0}^{\infty} \mathcal R_j 
\left( \frac{x}{2  i k}\right)^j, \quad x \to 0, 
\end{equation}
where
 $a = - \Gamma(2h)/\Gamma(h- i \qp)$, $b = \Gamma( 2-2h)/\Gamma(1-h-i\qp)  \times P$ and
$ \mathcal R_j := (h-i \qp)_j (1-h-i \qp)_j / j!$. The $k$-independent quantity  $P$ appearing above in $b$  
was previously given in Eq.~\eqref{eq:PtoQ ingoing}.

In general, the late-time asymptotics of
the Green function 
are determined by the singular points of $\tilde g$ in $\mathbb C$ with the largest real part \cite{Doetsch1974}. For 
perturbations of four-dimensional  stationary (asymptotically flat) extremal spacetimes, the known uppermost singular points of $\tilde g$ are the static $(\omega=0)$ and superradiant $(k=0)$ frequencies, both existing as branch points on the real axis. These branch points arise because the extremal radial equation has \emph{irregular} singular points at the horizon and infinity. We have found that the power-law contribution from the branch point at the static frequency is subleading in the case of extremal Kerr \cite{CGZsoon} and we assume the same to hold for extremal Reissner-Nordstr\"om.  

First consider modes having supplemental representations, where $h>1/2$. In this case the $a$-term in the denominator of $\tilde g$ in \eqref{eq:transfer} is more singular near $k = 0$. Then, the $n$-th radial derivative of the transfer function at the horizon to leading order in the late-time ``$1/k$'' expansion \cite{Casals:2016mel} 
 is given by
\begin{equation}\label{eq:tg real h}
 \tilde g^{(n)}_{ S } \vert_{\mathcal H} (k\to 0)  \sim f_n(x') ( - 2 i k)^{h-1-n+ i \qp}, \quad h >1/2,
\end{equation}
where $f_n(x') :=  n! (-1)^n \, \mathcal R_n \, R_{\rm up}(x') $ and the subscript $S$ means supplementary.  Using  theorem 37.1 of \cite{Doetsch1974}, the results of which are succinctly summarized in \cite{Casals:2016mel},   
we find the inverse transform to be
\begin{equation}\label{eq:gv real h}
  g^{(n)}_S\vert_{\mathcal H} (v\to \infty) \sim \frac{ f_n(x') e^{-i q v} }{2 \Gamma(1-h+n- i \qp) } \, \left( \frac{v}{2} \right)^{n-h- i \qp}.  
\end{equation}

For principal representation modes, both the $a$ and $b$ 
terms in the Wronksian (denominator of \eqref{eq:transfer}) are equally singular  \cite{Casals:2016mel} at $k=0$.  
The $n$-th derivative of the horizon transfer function in this case is
\begin{equation}\label{eq:tg complex h} 
  \tilde g^{(n)}_P \vert_{\mathcal H} (k\to 0) \sim \frac{\mathfrak{f}_n(x') (-2 i k)^{\alpha }}{ (-2 i k)^{-2 i \lambda} +\zeta },  \quad h \in \mathbb{C},
\end{equation}
where we have introduced
$ \mathfrak{f}_n(x') :=   (-1)^n n! \mathcal R_n\, R_{\rm up}(x') /a$,
 $ \zeta :=  b/a$ , and  $\quad \alpha := - 1/2 - n+  i(\qp- \lambda)$ as simplifying factors.  Despite considerable effort, we are unable to analytically invert  \eqref{eq:tg complex h}. Instead, we resort to numerical evaluation with the \emph{Mathematica} package \textsf{NumericalLaplaceInversion} \cite{VALKO2004629}, which computes the inverse Laplace transform to arbitrary numerical precision. Through previous experience with the same form of integral  \cite{Casals:2016mel}, we find it may be fitted to 
\begin{equation}\label{eq:the integral defeats all}
g^{(n)}_P\vert_{\mathcal H} (v\to \infty)  \sim D_n e^{-i q v}  v^{n-1/2 + ip},
\end{equation} 
 for complex $D_n$ and $p \in \mathbb R$.  Comparison with \eqref{eq:gv real h} reveals that the principal modes are dominant at late times.

\section{Symmetry Interpretation}\label{sec:symmetry}
We have shown that a collective ringing of overtones at the critical frequency gives rise to transient power-law behavior with duration $1/\sigma$. Though the field itself decays, the $n$-th transverse derivative grows with scaling $\sigma^{1/2-n}$. Therefore, the energy density $T_{\alpha \beta} u^\alpha u^\beta \sim \sigma^{-1}$ measured by infalling observers at the horizon becomes infinite in the extremal limit.

The same phenomena were found for nonaxisymmetric perturbations of the near-extremal and extremal Kerr black holes \cite{Casals:2016mel, Gralla:2016qfw}.  In hindsight, the similarities may have been predicted from their mutual 
symmetries. Both scenarios contain a $U(1)$  symmetry in addition to the near-horizon 2+1 Lorentz symmetry $SO(2,1)$. For Kerr the $U(1)$ manifests as axisymmetry of spacetime, while for charged scalars 
 it is a gauge symmetry on the principal bundle with $q$ playing the role of $\partial_\phi$ angular momentum $m$.  The triviality of the $U(1)$ bundle---existence of a global section---enabled us to globally utilize a smooth gauge adapted to the $\lambda \to 0$ limit wherein the connection between $m$ and $q$ was furnished 
 explicitly through the near-horizon frequency $\bar \omega = r_+(\omega - q \Phi_+)/\lambda$. In this way, the nonaxisymmetric  modes of rapidly spinning Kerr may be thought of as a tower of charged scalar perturbations with integer separated charges. 

We now sketch how the near-horizon charged perturbations of extremal Reissner-Nordstr\"om fall into modules 
of the symmetry group. 
 Symmetry in the background is defined as 
\begin{subequations}
\begin{align}
\mathcal L_{X,\chi} g_{\alpha\beta} &= \mathcal{L}_X g_{\alpha\beta} =0, \\
\mathcal L_{X,\chi} A &= \mathcal{L}_X A - d \chi =0,
\end{align}
\end{subequations}
where $X$ is the set of left-invariant vector fields generating spacetime isometries  and $\chi$ is the set of $U(1)$ transition functions, which, for smooth $A$, can be uniquely paired with each $X$ \cite{Prabhu:2015vua}.  The generators act on the charged scalar by $\mathcal L_{X,\chi} \psi  = \mathcal L_{X} \psi + i q \chi \psi.$ 
In the extremal limit the $\mathfrak{so}(2,1)$  Cartan-Weyl generators 
carry simple expressions in ingoing coordinates 
\begin{subequations}
\begin{align}
H_0 &=  v \partial_v - x \partial_x,  \\
 H_+ &= \pd_v,  \\
  H_- &= v^2 \partial_v - 2 (xv+1)\partial_x,
\end{align}
\end{subequations}
and satisfy the commutation relations $ [ H_+,H_-] = 2 H_0$ and $[H_{\pm},H_0]=H_{\pm}.$
These vector fields Lie derive the metric \eqref{eq:NHERN}. 
To Lie derive the vector potential
 $A = v dx$
 we pair $H_-$ with the transition function
 $\chi_- = -2v$. We also pair $H_0$ with $\chi_0=1$.  With this pairing all generators commute with the $D^2$ operator. Then, by Schur's lemma, $D^2$ is proportional to the identity.   Therefore   
 we have
\begin{align}\label{eq:casimir}
 D^2 \psi = 
 \Big (h(h-1) - K_{\ell} + C \Big) \psi
\end{align}
where $K_\ell$ labels  $\mathfrak{so}(3)$ as before, 
$C=\qp^2$ trivially labels $\mathfrak{u}(1)$, and 
 $h(h-1)$ is the   $\mathfrak{so}(2,1)$ Casimir 
 \cite{Wybourne1974book}.  The wave equation then implies 
$h = \frac12 \pm \sqrt{1/4+ K_\ell-C}$ for nontrivial solutions.  
Thus for a given $\ell$ and $q$, the near-horizon solutions $\psi$ may be partially classified according to $h$. 
The Casimir $h$ provides only a partial classification because $\mathfrak{ so}(2,1)$ is noncompact. In general, simple noncompact Lie algebras are classified not only by the Casimir $h(h-1)$, but also by a set of roots---simultaneous eigenvalues of the diagonal generators. 
For $\mathfrak{ so}(2,1)$, the Cartan subalgebra is rank 1 (single eigenvalue), so the choice of generator we diagonalize is arbitrary without restricting to unitary representations. Our choice is to classify by the eigenvalue $E_0$ of $H_0$, which is already diagonal in the ingoing coordinate basis, 
\[ \mathcal L_{H_0} \psi = E_0 \psi .\]  One solution of this equation is $\psi = v^{E_0} f(vx)$, while another is $\psi = x^{-E_0} f(vx)$. 
The physical eigensolution is determined by fixing boundary conditions such as regularity at $x=0$. 

General finite representations of $SO(2,1)$ are classified into four types, three of which have unbounded weight spectra $E_0 + n$, where $n \in \mathbb Z$, and the other is finite 
 \cite{Bargmann1947, Barut:1965, Gitman1997, Balasubramanian:1998sn, basu1982}.  The infinite representations are further divided into principal/supplementary series which have a bilateral unbounded weight spectrum,  and highest/lowest weight series which are unbounded from above/below $n=0$, respectively.  For the principal/supplementary series modes, $h$ is complex/real, whereas  the highest and lowest weight modes have real $h=\pm E_0$. For the finite representation, which occurs for $\qp=0$, $h$ is an integer. The only unitary finite representation is trivial ($\ell=0$).  Axisymmetric linear perturbations of Kerr transform as a finite representation also. 

We have been primarily interested in the principal series representations, which have the dominant contribution to the horizon instability.
As $K_{\ell}$ is positive semidefinite, the field is a principal series solution of \eqref{eq:casimir}   when  
\begin{equation}\label{eq:C eq}
C > 1/4 + K_\ell.
\end{equation}
 For charged fields, $C$ is an arbitrary parameter that may  be chosen such that Eq.~\eqref{eq:C eq} is satisfied for some $\ell$. In the case of generic mode perturbations of Kerr
\footnote{For scalar fields in Kerr  $K = A_{\ell m \omega} + a^2 \omega^2 $ where $A_{\ell m \omega}$ is the spheroidal eigenvalue. It can be shown that $K \geq \abs m ( \abs m + 1)$. },
 principal representations correspond to $1/4 \lesssim m/\ell$ \cite{bardeen-horowitz1999}.  
In near-horizon regions of both RN and Kerr, the principal harmonics (modes on which the generators act such that \eqref{eq:casimir} and \eqref{eq:C eq} are satisfied)
 which respect \emph{causal} boundary conditions, and continuously match to the far region radiation, realize  nonunitary representations.  With respect to the global near-horizon $\rm{ AdS}_2$ geometry, these  modes  lack a positive definite Hermitian form \cite{Kunze1960} and violate the Breitenlohner-Freedman bound \cite{Breitenlohner:1982jf}.  Stability results supporting these observations were found for the near-horizon perturbations of extremal Kerr in \cite{Dias:2009ex,Amsel:2009ev}. 

 Lastly we remark on the explicit action of $\mathfrak {so}(2,1) \in \mathfrak{g}$ on the physical solutions. With the $v$ and $x$ dependence derived in \eqref{eq:Searly}, let us write $\psi = v^{h-1-i\qp} f( vx) $, ignoring overall constants and also the angular dependence. As $f(vx)$ is trivially Lie derived along $H_0$,  $\mathcal L_{H_0} \psi = (h-1) \psi$ follows \footnote{The $ -i \qp$ in $E_0$ was removed by automorphically shifting $H_0$ by the $\mathfrak u(1)$ generator.}. This weight spectrum is unbounded, $(\mathcal L_{H_{\pm}} )^j \psi \neq 0 \, \, \forall j \in \mathbb N$, and has neither a highest nor lowest weight module. Starting with $\psi$ one can generate descendant solutions of \eqref{eq:casimir} in both stable and unstable directions by acting on $\psi$ with either $H_+$ or $H_-$  at the horizon: $H_-$ raises $E_0$ by an integer and enhances the Aretakis rate by an integer power of $v$, while $H_+$ lowers it in the same way in the opposite direction. Though we can map solutions into solutions by the action of $\mathfrak{so}(2,1)$, to maintain consistency with the boundary conditions, one must also act with the Cartan-Weyl generators of $\mathfrak{so}(3)$, changing the multipole number, \emph{and} adjust $q$ suitably to ensure uniqueness of the solution. This ``mixing'' limits the practical application of  the construction based on symmetry alone. 

\section*{Acknowledgements}
We thank Sam Gralla, Stephen Green, Stefan Hollands, Drew Jamieson, Kartik Prabhu, and Leo Stein  for encouragement and helpful conversations. 

This work was supported in part by NSF Grant PHY--1506027 to the University of Arizona. Hospitality during the beginning stages of this work was provided by the Perimeter Institute for Theoretical Physics. Research at Perimeter Institute is supported by the Government of Canada through Industry Canada and by the Province of Ontario through the Ministry of Economic Development \& Innovation.

\section{appendix}
\subsection{Combined symmetries}
 We work in the near-horizon region, which is diffeomorphic at constant angular coordinates to $\mathrm{AdS}_{2}$ in the Poincar\'e  patch
\begin{equation}\label{eq:Ads2PP}
	ds^{2} = -\brx (\brx+1)d\brt^2 + \frac{d\brx^2 }{\brx (\brx+1)}.
\end{equation}
The Killing vector fields of \eqref{eq:Ads2PP}
 are given by
\begin{align}\label{eq:generators}
H_0 & = -2\pd_{\brt} \nonumber,\\
 \qquad{H}_\pm &= \frac{e^{\pm\brt/2}}{\sqrt{\brx(\brx+1)}} \Big( (2\brx+1)\pd_{\brt} \mp\brx(\brx+1)\pd_{\brx} \Big).
\end{align}
They satisfy $\mathfrak{sl}_2(\mathbb R)$ commutation relations $[  H_+,  H_-] = 2 H_0$ and $[ H_{\pm}, H_0]=  H_{\pm}$.  In our adopted gauge, where $A=\brx d\brt$, the potential $A$ is Lie-derived by $ H_0$, but fails to be symmetric under $ H_-$ and $ H_+$.  In fact, there is no gauge where $ A$ is invariant under all three generators.  

We say that there is a combined symmetry of the metric $g$ and electromagnetic potential $A$ 
if there is a vector $X$ such that $\Lie_X g=0$ and $\Lie_X A = d f$ for some function $f$.
A suitable notion of Lie derivative for the combination is given by
\begin{subequations}\label{eq:Lie2}
\begin{align}
\bar{\Lie}_{\bar{X}} {g}_{ab} & = \Lie_X {g}_{ab}, \\
\bar{\Lie}_{\bar{X}} {A} & = \Lie_X {A} +  d\zeta,\\
\bar{\Lie}_{\bar{X}} {\psi} & = \Lie_X {\psi} + i q \zeta {\psi}. \label{eq:Lie3}
\end{align}
\end{subequations}
where 
\begin{align}
    \bar{X} = (X,\zeta).
\end{align}
We make the choice of pairings
\begin{align}\label{eq:symm pairs}
   \bar{H}_0 = (H_0, \zeta_0), \quad \bar{H}_+ = (  H_+, \zeta_+),  \quad \bar{H}_- = (  H_-,\zeta_-),
\end{align}
 by combining \eqref{eq:generators} with the three scalars
\begin{equation}\label{eq:zetas}
\zeta_0=-r_+, \qquad \zeta_\pm=\frac{r_+ e^{\pm\brt/2}}{\sqrt{\brx(\brx+1)}}. 
\end{equation}
  Under a pure $U(1)$ transformation, $A \to A + d \Lambda$, and therefore  $\bar{X}=(X,\zeta)$ changes by $\zeta \to \zeta - \Lie_X \Lambda$.  For highly spinning axisymmetric black holes, the $\phi$-symmetry can be characterized  in the limiting spacetime near the horizon in a similar fashion to this gauge symmetry. The associated quantum numbers in the Kerr case are effective charges for each azimuthal quantum number \cite{Dias:2009ex}.
  
\subsection{QNM wavefunctions }\label{sec:modules}

The generators act on modes\footnote{Here, for convenience, we choose ``Laplace'' vs. ``Fourier'' to represent our modes. The Laplace frequencies $s_{n}$ are related to the Fourier frequencies $\bar{k}_{n}$ used in the text
by $\bar{k}_{n}=-2i\brs_{n}$. The formula for a QNM written here is for a supplementary mode (see text for definition). Its principal partner mode can be obtained
via $h\to1-h$}
\begin{equation}\label{eq:phi n}
\phi_n(\brt,\brx) = b_n(\brx) e^{\brs_n \brt}, \qquad \brs_{s} = -(h+n+i\qp)/2,
\end{equation}
where
\begin{align}
b_n &= \left(-\brx\right)^{\brs_n}
 \left(1+\brx\right)^{- \brs_n - i \qp } \nonumber \\
&\times  {}_2F_1(1-h-i\qp,h-i\qp,1-h-n-i\qp,-\brx)\label{eq:bnear1}\\
  &= \left(-\brx\right)^{\brs_n} \left(1+\brx\right)^{ \brs_n + i \qp }
   \sum_{j=0}^n P^{(n)}_j \left(-\brx\right)^j,
\end{align}
  as scalar operators to give
\begin{align}\label{eq:operator actions}
	\bar\Lie_{\bar H_0}\phi_n &=\left(h+n\right)\phi_n, \nonumber \\
	\bar\Lie_{\bar H_+}\phi_n &=\alpha^+_n \phi_{n-1}, \nonumber \\
	\bar\Lie_{\bar H_-}\phi_n &=\alpha^-_n\phi_{n+1}, 
\end{align}
where
\begin{equation}\label{eq:alphas}
\alpha_n^{+}=\frac{in(n+2h-1)}{n+h-1+i\qp}, \qquad \alpha_n^{-}=-i(h+n+i\qp).
\end{equation}
The Casimir operator 
\begin{equation}\label{eq:Cas}
	\Omega= \bar\Lie_{\bar{H}_0} (\bar\Lie_{\bar{H}_0}- 1) - \bar\Lie_{\bar{H}_-} \bar\Lie_{\bar{H}_+}
\end{equation}
 acts on QNM wavefunctions $\phi_{n}$ in the usual way
\begin{equation}
\Omega \phi_n=h(h-1)\phi_n.
\end{equation}
Note that the eigenvalue of $\bar H_0$ is $h$ modulo a non-negative integer
and that the $n=0$ mode is annihilated by $\bar H_+$. By acting with $\bar H_-$ recursively on $\phi_{0}$, one can construct a ``state'' with arbitrary overtone number $n$.

We finish by calculating the normalization of the modes. For this, we use a canonical framework, utilizing the map
\begin{align}\label{eq:Qmap}
\mathcal Q \phi &: \phi \to \boldsymbol{\phi} \nonumber ,\\
 \boldsymbol{\phi} &= \begin{pmatrix} \phi \\ \pi \end{pmatrix}=\begin{pmatrix} \phi \\  (\pd_{\brt}-iqA_{\brt}) {\phi} \end{pmatrix}.
\end{align}
Under this map 
\begin{align}
\mathcal Q(H_0 \phi) &= -2  
  \begin{pmatrix} 
    iqA_{\brt} & 1 \\
    L & iqA_{\brt} 
  \end{pmatrix} \begin{pmatrix} \phi \\ \pi \end{pmatrix} \nonumber \\
  	&=-2{\boldsymbol{\mathcal H}} \boldsymbol{\phi},
\end{align}
where $L=\brx (1+\brx)\pd_{\brx} \left( \brx\left(1+\brx\right)\right)\pd_{\brx} $ and ${\boldsymbol{\mathcal H}}$ is the analogue of a Hamiltonian matrix 
\begin{align}
	\boldsymbol{\mathcal H}\boldsymbol{\phi}_n=s_n\boldsymbol{\phi}_n.
\end{align}
For the shifted generator $\bar{H}_{0}$, we get simply
$\mathcal Q( \bar H_0 \phi ) = (-2\boldsymbol{\mathcal H}-i\qp\boldsymbol{1})\boldsymbol{\phi}$,
where $\boldsymbol{1}$ is the $2\times 2$ identity matrix.
After a straightforward computation, one finds
\begin{widetext}
\begin{align} 
\mathcal Q(\bar H_\pm \phi) &= \frac{e^{\pm \brt/2}}{\sqrt{\brx(1+\brx)}}
 \begin{pmatrix} 
   2 i\qp \brx(1+\brx) \mp \brx(1+\brx)\partial_{\brx} & 1+2\brx \\
 (1+2\brx) L -\tfrac12x(1+x)\partial_{\brx} & \pm\tfrac{1}{2}+ \brx(\pm1+2i\qp(1+\brx))\mp \brx(1+\brx)\pd_x
  \end{pmatrix} \begin{pmatrix} \phi \\ \pi \end{pmatrix}.
\end{align}
\end{widetext}
A symmetric bilinear  $( \boldsymbol\phi_{n},\boldsymbol \phi_{m})$ can be obtained as in \cite{greenumeey} from the symplectic form  $W(\boldsymbol \phi_{1},\boldsymbol \phi_{2}) = \int_{\Sigma} \pi_{1}\phi_{2}-\pi_{2}\phi_{1}$ by subjecting one of the fields to a combined charge conjugation and time reversal $( \boldsymbol\phi_{n},\boldsymbol \phi_{n}) \equiv W(\mathsf{CT}\boldsymbol \phi_{n},\boldsymbol \phi_{m})$. After a calculation, this gives
\begin{widetext}
\begin{align}\label{eq:near-near R bilinear}
(\boldsymbol{\phi}_n,\boldsymbol{\phi}_m) =  \  &\frac{(-1)^{-\brs_n-\brs_m}}{\Gamma(1-2i\qp-\brs_n-\brs_m)} 
\Bigg( \left(\brs_n+\brs_m\right)\sum_{k=0}^{n+m}(-1)^kP_k^{(n,m)}\Gamma(1-k-2i\qp-2\brs_n-2\brs_m)\Gamma(k+\brs_n+\brs_m) 
\nonumber\\
&\quad \quad \quad \quad \quad \quad\quad \quad   -2i\qp\sum_{k=0}^{n+m}(-1)^kP_k^{(n,m)}
\Gamma(-k-2i\qp-2\brs_n-2\brs_m)\Gamma(k+1+\brs_n+\brs_m) 
\Bigg),
\end{align}
\end{widetext}
where $P_{j}^{(n)}=(-n)_{j}(1-2h-n)_{j}/(j!(1-h-n-i\qp)_{j})$ and $P_{j}^{(n,m)}=\sum_{k=0}^{j}P^{(n)}_{j-k}P^{(m)}_{k}$. 
With respect to \eqref{eq:near-near R bilinear} the operators \eqref{eq:symm pairs} have the symmetries
\begin{align}\label{eq: adjoints}
 	(\boldsymbol{\phi}_n, \bar H_0 \boldsymbol \phi_m) &= (\bar H_0 \boldsymbol \phi_n, \boldsymbol \phi_m),\nonumber \\
	 (\boldsymbol{\phi}_n, \bar H_\pm \boldsymbol \phi_m) &= (\bar H_\mp \boldsymbol \phi_n, \boldsymbol \phi_m).
\end{align}

To compute the matrix elements, one introduces normalized modes
\begin{equation}\label{eq:normalized modes}
\hat{\bs\phi}_n= N_n^{-1/2} \bs\phi_n,
\end{equation}
 where the normalization 
\begin{align}\label{eq:NHM N}
N_{n} = \frac{(-1)^{h+i\qp}}{\Gamma(h-i\qp)}
\frac{2n!\Gamma(2h+n)\Gamma(1-h-n-i\qp)}{(h+i\qp)_n}
\end{align}
is chosen such that 
\begin{equation}
	(\hat{\bs\phi}_n,\hat{\bs\phi}_m)=\delta_{nm}.
\end{equation}
Lastly, we give a recursion relation for the coefficients $\alpha^{\pm}_{n}$ in terms of the constants $N_n$.  From Eq.~\eqref{eq: adjoints} it follows that
\begin{align*}
	(\bar H_+\hat{\bs \phi}_{n}, \hat{\bs \phi}_{n-1})=(\hat{\bs \phi}_{n}, \bar H_-\hat{\bs \phi}_{n-1}).
\end{align*}
This, together with application of Eqs~\eqref{eq:normalized modes}, \eqref{eq:operator actions}, and \eqref{eq:alphas} gives the relation  
\begin{equation}
	\frac{N_{n}}{N_{n-1}}=
	\frac{\alpha_n^+}{\alpha_{n-1}^-}=\frac{n(1-2h-n)}{(n+h-1+i\qp)^2}
\end{equation} 
among the constants.

\bibliographystyle{apsrev4-1}
\bibliography{/Users/peter/Documents/MyReferences.bib}

\begin{thebibliography}{82}%
\makeatletter
\providecommand \@ifxundefined [1]{%
 \@ifx{#1\undefined}
}%
\providecommand \@ifnum [1]{%
 \ifnum #1\expandafter \@firstoftwo
 \else \expandafter \@secondoftwo
 \fi
}%
\providecommand \@ifx [1]{%
 \ifx #1\expandafter \@firstoftwo
 \else \expandafter \@secondoftwo
 \fi
}%
\providecommand \natexlab [1]{#1}%
\providecommand \enquote  [1]{``#1''}%
\providecommand \bibnamefont  [1]{#1}%
\providecommand \bibfnamefont [1]{#1}%
\providecommand \citenamefont [1]{#1}%
\providecommand \href@noop [0]{\@secondoftwo}%
\providecommand \href [0]{\begingroup \@sanitize@url \@href}%
\providecommand \@href[1]{\@@startlink{#1}\@@href}%
\providecommand \@@href[1]{\endgroup#1\@@endlink}%
\providecommand \@sanitize@url [0]{\catcode `\\12\catcode `\$12\catcode
  `\&12\catcode `\#12\catcode `\^12\catcode `\_12\catcode `\%12\relax}%
\providecommand \@@startlink[1]{}%
\providecommand \@@endlink[0]{}%
\providecommand \url  [0]{\begingroup\@sanitize@url \@url }%
\providecommand \@url [1]{\endgroup\@href {#1}{\urlprefix }}%
\providecommand \urlprefix  [0]{URL }%
\providecommand \Eprint [0]{\href }%
\providecommand \doibase [0]{http://dx.doi.org/}%
\providecommand \selectlanguage [0]{\@gobble}%
\providecommand \bibinfo  [0]{\@secondoftwo}%
\providecommand \bibfield  [0]{\@secondoftwo}%
\providecommand \translation [1]{[#1]}%
\providecommand \BibitemOpen [0]{}%
\providecommand \bibitemStop [0]{}%
\providecommand \bibitemNoStop [0]{.\EOS\space}%
\providecommand \EOS [0]{\spacefactor3000\relax}%
\providecommand \BibitemShut  [1]{\csname bibitem#1\endcsname}%
\let\auto@bib@innerbib\@empty
\bibitem [{\citenamefont {Oppenheimer}\ and\ \citenamefont
  {Snyder}(1939)}]{OppenheimerSnyder:1939}%
  \BibitemOpen
  \bibfield  {author} {\bibinfo {author} {\bibfnamefont {J.~R.}\ \bibnamefont
  {Oppenheimer}}\ and\ \bibinfo {author} {\bibfnamefont {H.}~\bibnamefont
  {Snyder}},\ }\href {\doibase 10.1103/PhysRev.56.455} {\bibfield  {journal}
  {\bibinfo  {journal} {Phys. Rev.}\ }\textbf {\bibinfo {volume} {56}},\
  \bibinfo {pages} {455} (\bibinfo {year} {1939})}\BibitemShut {NoStop}%
\bibitem [{\citenamefont {{Blandford}}\ and\ \citenamefont
  {{Znajek}}(1977)}]{BlandfordZnajek:1977}%
  \BibitemOpen
  \bibfield  {author} {\bibinfo {author} {\bibfnamefont {R.~D.}\ \bibnamefont
  {{Blandford}}}\ and\ \bibinfo {author} {\bibfnamefont {R.~L.}\ \bibnamefont
  {{Znajek}}},\ }\href {\doibase 10.1093/mnras/179.3.433} {\bibfield  {journal}
  {\bibinfo  {journal} {\mnras}\ }\textbf {\bibinfo {volume} {179}},\ \bibinfo
  {pages} {433} (\bibinfo {year} {1977})}\BibitemShut {NoStop}%
\bibitem [{\citenamefont {Abbott}\ \emph {et~al.}(2016)\citenamefont {Abbott}
  \emph {et~al.}}]{TheLIGOScientific:2016qqj}%
  \BibitemOpen
  \bibfield  {author} {\bibinfo {author} {\bibfnamefont {B.~P.}\ \bibnamefont
  {Abbott}} \emph {et~al.} (\bibinfo {collaboration} {Virgo, LIGO
  Scientific}),\ }\href@noop {} {\  (\bibinfo {year} {2016})},\ \Eprint
  {http://arxiv.org/abs/1602.03839} {arXiv:1602.03839 [gr-qc]} \BibitemShut
  {NoStop}%
\bibitem [{\citenamefont {Bekenstein}(1973)}]{Bekenstein:1973ur}%
  \BibitemOpen
  \bibfield  {author} {\bibinfo {author} {\bibfnamefont {J.~D.}\ \bibnamefont
  {Bekenstein}},\ }\href {\doibase 10.1103/PhysRevD.7.2333} {\bibfield
  {journal} {\bibinfo  {journal} {Phys. Rev.}\ }\textbf {\bibinfo {volume}
  {D7}},\ \bibinfo {pages} {2333} (\bibinfo {year} {1973})}\BibitemShut
  {NoStop}%
\bibitem [{\citenamefont {Hawking}(1975)}]{Hawking:1974sw}%
  \BibitemOpen
  \bibfield  {author} {\bibinfo {author} {\bibfnamefont {S.~W.}\ \bibnamefont
  {Hawking}},\ }\bibfield  {booktitle} {\emph {\bibinfo {booktitle} {{In
  *Gibbons, G.W. (ed.), Hawking, S.W. (ed.): Euclidean quantum gravity*
  167-188}}},\ }\href {\doibase 10.1007/BF02345020} {\bibfield  {journal}
  {\bibinfo  {journal} {Commun. Math. Phys.}\ }\textbf {\bibinfo {volume}
  {43}},\ \bibinfo {pages} {199} (\bibinfo {year} {1975})},\ \bibinfo {note}
  {[,167(1975)]}\BibitemShut {NoStop}%
\bibitem [{\citenamefont {Dafermos}\ \emph {et~al.}(2016)\citenamefont
  {Dafermos}, \citenamefont {Holzegel},\ and\ \citenamefont
  {Rodnianski}}]{Dafermos:2016uzj}%
  \BibitemOpen
  \bibfield  {author} {\bibinfo {author} {\bibfnamefont {M.}~\bibnamefont
  {Dafermos}}, \bibinfo {author} {\bibfnamefont {G.}~\bibnamefont {Holzegel}},
  \ and\ \bibinfo {author} {\bibfnamefont {I.}~\bibnamefont {Rodnianski}},\
  }\href@noop {} {\  (\bibinfo {year} {2016})},\ \Eprint
  {http://arxiv.org/abs/1601.06467} {arXiv:1601.06467 [gr-qc]} \BibitemShut
  {NoStop}%
\bibitem [{\citenamefont {Dafermos}\ \emph {et~al.}(2014)\citenamefont
  {Dafermos}, \citenamefont {Rodnianski},\ and\ \citenamefont
  {Shlapentokh-Rothman}}]{Dafermos:2014cua}%
  \BibitemOpen
  \bibfield  {author} {\bibinfo {author} {\bibfnamefont {M.}~\bibnamefont
  {Dafermos}}, \bibinfo {author} {\bibfnamefont {I.}~\bibnamefont
  {Rodnianski}}, \ and\ \bibinfo {author} {\bibfnamefont {Y.}~\bibnamefont
  {Shlapentokh-Rothman}},\ }\href@noop {} {\  (\bibinfo {year} {2014})},\
  \Eprint {http://arxiv.org/abs/1402.7034} {arXiv:1402.7034 [gr-qc]}
  \BibitemShut {NoStop}%
\bibitem [{\citenamefont {Finster}\ and\ \citenamefont
  {Smoller}(2016)}]{Finster:2016tky}%
  \BibitemOpen
  \bibfield  {author} {\bibinfo {author} {\bibfnamefont {F.}~\bibnamefont
  {Finster}}\ and\ \bibinfo {author} {\bibfnamefont {J.}~\bibnamefont
  {Smoller}},\ }\href@noop {} {\  (\bibinfo {year} {2016})},\ \Eprint
  {http://arxiv.org/abs/1606.08005} {arXiv:1606.08005 [math-ph]} \BibitemShut
  {NoStop}%
\bibitem [{\citenamefont {{Strominger}}\ and\ \citenamefont
  {{Vafa}}(1996)}]{Strominger1996}%
  \BibitemOpen
  \bibfield  {author} {\bibinfo {author} {\bibfnamefont {A.}~\bibnamefont
  {{Strominger}}}\ and\ \bibinfo {author} {\bibfnamefont {C.}~\bibnamefont
  {{Vafa}}},\ }\href {\doibase 10.1016/0370-2693(96)00345-0} {\bibfield
  {journal} {\bibinfo  {journal} {Physics Letters B}\ }\textbf {\bibinfo
  {volume} {379}},\ \bibinfo {pages} {99} (\bibinfo {year} {1996})},\ \Eprint
  {http://arxiv.org/abs/arXiv:hep-th/9601029} {arXiv:hep-th/9601029}
  \BibitemShut {NoStop}%
\bibitem [{\citenamefont {{Teukolsky}}\ and\ \citenamefont
  {{Press}}(1974)}]{TeukolskyPress1974}%
  \BibitemOpen
  \bibfield  {author} {\bibinfo {author} {\bibfnamefont {S.~A.}\ \bibnamefont
  {{Teukolsky}}}\ and\ \bibinfo {author} {\bibfnamefont {W.~H.}\ \bibnamefont
  {{Press}}},\ }\href {\doibase 10.1086/153180} {\bibfield  {journal} {\bibinfo
   {journal} {\apj}\ }\textbf {\bibinfo {volume} {193}},\ \bibinfo {pages}
  {443} (\bibinfo {year} {1974})}\BibitemShut {NoStop}%
\bibitem [{\citenamefont {{Porfyriadis}}\ and\ \citenamefont
  {{Strominger}}(2014)}]{porfyriadis-strominger2014}%
  \BibitemOpen
  \bibfield  {author} {\bibinfo {author} {\bibfnamefont {A.~P.}\ \bibnamefont
  {{Porfyriadis}}}\ and\ \bibinfo {author} {\bibfnamefont {A.}~\bibnamefont
  {{Strominger}}},\ }\href {\doibase 10.1103/PhysRevD.90.044038} {\bibfield
  {journal} {\bibinfo  {journal} {\prd}\ }\textbf {\bibinfo {volume} {90}},\
  \bibinfo {eid} {044038} (\bibinfo {year} {2014})},\ \Eprint
  {http://arxiv.org/abs/1401.3746} {arXiv:1401.3746 [hep-th]} \BibitemShut
  {NoStop}%
\bibitem [{\citenamefont {{Hadar}}\ \emph {et~al.}(2014)\citenamefont
  {{Hadar}}, \citenamefont {{Porfyriadis}},\ and\ \citenamefont
  {{Strominger}}}]{hadar-porfyriadis-strominger2014}%
  \BibitemOpen
  \bibfield  {author} {\bibinfo {author} {\bibfnamefont {S.}~\bibnamefont
  {{Hadar}}}, \bibinfo {author} {\bibfnamefont {A.~P.}\ \bibnamefont
  {{Porfyriadis}}}, \ and\ \bibinfo {author} {\bibfnamefont {A.}~\bibnamefont
  {{Strominger}}},\ }\href {\doibase 10.1103/PhysRevD.90.064045} {\bibfield
  {journal} {\bibinfo  {journal} {\prd}\ }\textbf {\bibinfo {volume} {90}},\
  \bibinfo {eid} {064045} (\bibinfo {year} {2014})},\ \Eprint
  {http://arxiv.org/abs/1403.2797} {arXiv:1403.2797 [hep-th]} \BibitemShut
  {NoStop}%
\bibitem [{\citenamefont {{Lupsasca}}\ \emph {et~al.}(2014)\citenamefont
  {{Lupsasca}}, \citenamefont {{Rodriguez}},\ and\ \citenamefont
  {{Strominger}}}]{lupsasca-rodriguez-strominger2014}%
  \BibitemOpen
  \bibfield  {author} {\bibinfo {author} {\bibfnamefont {A.}~\bibnamefont
  {{Lupsasca}}}, \bibinfo {author} {\bibfnamefont {M.~J.}\ \bibnamefont
  {{Rodriguez}}}, \ and\ \bibinfo {author} {\bibfnamefont {A.}~\bibnamefont
  {{Strominger}}},\ }\href {\doibase 10.1007/JHEP12(2014)185} {\bibfield
  {journal} {\bibinfo  {journal} {J.~High~Energy~Phys.}\ }\textbf {\bibinfo
  {volume} {12}},\ \bibinfo {eid} {185} (\bibinfo {year} {2014})},\ \Eprint
  {http://arxiv.org/abs/1406.4133} {arXiv:1406.4133 [hep-th]} \BibitemShut
  {NoStop}%
\bibitem [{\citenamefont {{Zhang}}\ \emph {et~al.}(2014)\citenamefont
  {{Zhang}}, \citenamefont {{Yang}},\ and\ \citenamefont
  {{Lehner}}}]{zhang-yang-lehner2014}%
  \BibitemOpen
  \bibfield  {author} {\bibinfo {author} {\bibfnamefont {F.}~\bibnamefont
  {{Zhang}}}, \bibinfo {author} {\bibfnamefont {H.}~\bibnamefont {{Yang}}}, \
  and\ \bibinfo {author} {\bibfnamefont {L.}~\bibnamefont {{Lehner}}},\ }\href
  {\doibase 10.1103/PhysRevD.90.124009} {\bibfield  {journal} {\bibinfo
  {journal} {\prd}\ }\textbf {\bibinfo {volume} {90}},\ \bibinfo {eid} {124009}
  (\bibinfo {year} {2014})},\ \Eprint {http://arxiv.org/abs/1409.0345}
  {arXiv:1409.0345 [astro-ph.HE]} \BibitemShut {NoStop}%
\bibitem [{\citenamefont {{Lupsasca}}\ and\ \citenamefont
  {{Rodriguez}}(2015)}]{lupsasca-rodriguez2015}%
  \BibitemOpen
  \bibfield  {author} {\bibinfo {author} {\bibfnamefont {A.}~\bibnamefont
  {{Lupsasca}}}\ and\ \bibinfo {author} {\bibfnamefont {M.~J.}\ \bibnamefont
  {{Rodriguez}}},\ }\href {\doibase 10.1007/JHEP07(2015)090} {\bibfield
  {journal} {\bibinfo  {journal} {Journal of High Energy Physics}\ }\textbf
  {\bibinfo {volume} {7}},\ \bibinfo {eid} {90} (\bibinfo {year} {2015})},\
  \Eprint {http://arxiv.org/abs/1412.4124} {arXiv:1412.4124 [hep-th]}
  \BibitemShut {NoStop}%
\bibitem [{\citenamefont {{Hadar}}\ \emph {et~al.}(2015)\citenamefont
  {{Hadar}}, \citenamefont {{Porfyriadis}},\ and\ \citenamefont
  {{Strominger}}}]{hadar-porfyriadis-strominger2015}%
  \BibitemOpen
  \bibfield  {author} {\bibinfo {author} {\bibfnamefont {S.}~\bibnamefont
  {{Hadar}}}, \bibinfo {author} {\bibfnamefont {A.~P.}\ \bibnamefont
  {{Porfyriadis}}}, \ and\ \bibinfo {author} {\bibfnamefont {A.}~\bibnamefont
  {{Strominger}}},\ }\href {\doibase 10.1007/JHEP07(2015)078} {\bibfield
  {journal} {\bibinfo  {journal} {Journal of High Energy Physics}\ }\textbf
  {\bibinfo {volume} {7}},\ \bibinfo {eid} {78} (\bibinfo {year} {2015})},\
  \Eprint {http://arxiv.org/abs/1504.07650} {arXiv:1504.07650 [hep-th]}
  \BibitemShut {NoStop}%
\bibitem [{\citenamefont {{Gralla}}\ \emph {et~al.}(2015)\citenamefont
  {{Gralla}}, \citenamefont {{Porfyriadis}},\ and\ \citenamefont
  {{Warburton}}}]{GPW15}%
  \BibitemOpen
  \bibfield  {author} {\bibinfo {author} {\bibfnamefont {S.~E.}\ \bibnamefont
  {{Gralla}}}, \bibinfo {author} {\bibfnamefont {A.~P.}\ \bibnamefont
  {{Porfyriadis}}}, \ and\ \bibinfo {author} {\bibfnamefont {N.}~\bibnamefont
  {{Warburton}}},\ }\href {\doibase 10.1103/PhysRevD.92.064029} {\bibfield
  {journal} {\bibinfo  {journal} {\prd}\ }\textbf {\bibinfo {volume} {92}},\
  \bibinfo {eid} {064029} (\bibinfo {year} {2015})},\ \Eprint
  {http://arxiv.org/abs/1506.08496} {arXiv:1506.08496 [gr-qc]} \BibitemShut
  {NoStop}%
\bibitem [{\citenamefont {{Gralla}}\ \emph {et~al.}(2016)\citenamefont
  {{Gralla}}, \citenamefont {{Lupsasca}},\ and\ \citenamefont
  {{Strominger}}}]{gralla-lupsasca-strominger2016}%
  \BibitemOpen
  \bibfield  {author} {\bibinfo {author} {\bibfnamefont {S.~E.}\ \bibnamefont
  {{Gralla}}}, \bibinfo {author} {\bibfnamefont {A.}~\bibnamefont
  {{Lupsasca}}}, \ and\ \bibinfo {author} {\bibfnamefont {A.}~\bibnamefont
  {{Strominger}}},\ }\href {\doibase 10.1103/PhysRevD.93.104041} {\bibfield
  {journal} {\bibinfo  {journal} {\prd}\ }\textbf {\bibinfo {volume} {93}},\
  \bibinfo {eid} {104041} (\bibinfo {year} {2016})},\ \Eprint
  {http://arxiv.org/abs/1602.01833} {arXiv:1602.01833 [hep-th]} \BibitemShut
  {NoStop}%
\bibitem [{\citenamefont {{Comp{\`e}re}}\ and\ \citenamefont
  {{Oliveri}}(2016)}]{compere-oliveri2016}%
  \BibitemOpen
  \bibfield  {author} {\bibinfo {author} {\bibfnamefont {G.}~\bibnamefont
  {{Comp{\`e}re}}}\ and\ \bibinfo {author} {\bibfnamefont {R.}~\bibnamefont
  {{Oliveri}}},\ }\href {\doibase 10.1103/PhysRevD.93.024035} {\bibfield
  {journal} {\bibinfo  {journal} {\prd}\ }\textbf {\bibinfo {volume} {93}},\
  \bibinfo {eid} {024035} (\bibinfo {year} {2016})},\ \Eprint
  {http://arxiv.org/abs/1509.07637} {arXiv:1509.07637 [hep-th]} \BibitemShut
  {NoStop}%
\bibitem [{\citenamefont {{Porfyriadis}}\ \emph {et~al.}(2016)\citenamefont
  {{Porfyriadis}}, \citenamefont {{Shi}},\ and\ \citenamefont
  {{Strominger}}}]{porfyriadis-shi-strominger2016}%
  \BibitemOpen
  \bibfield  {author} {\bibinfo {author} {\bibfnamefont {A.~P.}\ \bibnamefont
  {{Porfyriadis}}}, \bibinfo {author} {\bibfnamefont {Y.}~\bibnamefont
  {{Shi}}}, \ and\ \bibinfo {author} {\bibfnamefont {A.}~\bibnamefont
  {{Strominger}}},\ }\href@noop {} {\bibfield  {journal} {\bibinfo  {journal}
  {ArXiv e-prints}\ } (\bibinfo {year} {2016})},\ \Eprint
  {http://arxiv.org/abs/1607.06028} {arXiv:1607.06028 [gr-qc]} \BibitemShut
  {NoStop}%
\bibitem [{\citenamefont {Yang}\ \emph {et~al.}(2015)\citenamefont {Yang},
  \citenamefont {Zimmerman},\ and\ \citenamefont {Lehner}}]{Yang:2014tla}%
  \BibitemOpen
  \bibfield  {author} {\bibinfo {author} {\bibfnamefont {H.}~\bibnamefont
  {Yang}}, \bibinfo {author} {\bibfnamefont {A.}~\bibnamefont {Zimmerman}}, \
  and\ \bibinfo {author} {\bibfnamefont {L.}~\bibnamefont {Lehner}},\ }\href
  {\doibase 10.1103/PhysRevLett.114.081101} {\bibfield  {journal} {\bibinfo
  {journal} {Phys. Rev. Lett.}\ }\textbf {\bibinfo {volume} {114}},\ \bibinfo
  {pages} {081101} (\bibinfo {year} {2015})},\ \Eprint
  {http://arxiv.org/abs/1402.4859} {arXiv:1402.4859 [gr-qc]} \BibitemShut
  {NoStop}%
\bibitem [{\citenamefont {{Cunningham}}\ and\ \citenamefont
  {{Bardeen}}(1972)}]{CunninghamBardeen1972}%
  \BibitemOpen
  \bibfield  {author} {\bibinfo {author} {\bibfnamefont {C.~T.}\ \bibnamefont
  {{Cunningham}}}\ and\ \bibinfo {author} {\bibfnamefont {J.~M.}\ \bibnamefont
  {{Bardeen}}},\ }\href {\doibase 10.1086/180933} {\bibfield  {journal}
  {\bibinfo  {journal} {\apjl}\ }\textbf {\bibinfo {volume} {173}},\ \bibinfo
  {pages} {L137} (\bibinfo {year} {1972})}\BibitemShut {NoStop}%
\bibitem [{\citenamefont {{Andersson}}\ and\ \citenamefont
  {{Glampedakis}}(2000)}]{Andersson2000}%
  \BibitemOpen
  \bibfield  {author} {\bibinfo {author} {\bibfnamefont {N.}~\bibnamefont
  {{Andersson}}}\ and\ \bibinfo {author} {\bibfnamefont {K.}~\bibnamefont
  {{Glampedakis}}},\ }\href {\doibase 10.1103/PhysRevLett.84.4537} {\bibfield
  {journal} {\bibinfo  {journal} {Physical Review Letters}\ }\textbf {\bibinfo
  {volume} {84}},\ \bibinfo {pages} {4537} (\bibinfo {year} {2000})},\ \Eprint
  {http://arxiv.org/abs/arXiv:gr-qc/9909050} {arXiv:gr-qc/9909050} \BibitemShut
  {NoStop}%
\bibitem [{\citenamefont {{Glampedakis}}\ and\ \citenamefont
  {{Andersson}}(2001)}]{Glampedakis2001}%
  \BibitemOpen
  \bibfield  {author} {\bibinfo {author} {\bibfnamefont {K.}~\bibnamefont
  {{Glampedakis}}}\ and\ \bibinfo {author} {\bibfnamefont {N.}~\bibnamefont
  {{Andersson}}},\ }\href {\doibase 10.1103/PhysRevD.64.104021} {\bibfield
  {journal} {\bibinfo  {journal} {\prd}\ }\textbf {\bibinfo {volume} {64}},\
  \bibinfo {pages} {104021} (\bibinfo {year} {2001})},\ \Eprint
  {http://arxiv.org/abs/arXiv:gr-qc/0103054} {arXiv:gr-qc/0103054} \BibitemShut
  {NoStop}%
\bibitem [{\citenamefont {Yang}\ \emph
  {et~al.}(2013{\natexlab{a}})\citenamefont {Yang}, \citenamefont {Zimmerman},
  \citenamefont {Zengino{\u g}lu}, \citenamefont {Zhang}, \citenamefont {Berti}
  \emph {et~al.}}]{Yang:2013uba}%
  \BibitemOpen
  \bibfield  {author} {\bibinfo {author} {\bibfnamefont {H.}~\bibnamefont
  {Yang}}, \bibinfo {author} {\bibfnamefont {A.}~\bibnamefont {Zimmerman}},
  \bibinfo {author} {\bibfnamefont {A.}~\bibnamefont {Zengino{\u g}lu}},
  \bibinfo {author} {\bibfnamefont {F.}~\bibnamefont {Zhang}}, \bibinfo
  {author} {\bibfnamefont {E.}~\bibnamefont {Berti}},  \emph {et~al.},\ }\href
  {\doibase 10.1103/PhysRevD.88.044047} {\bibfield  {journal} {\bibinfo
  {journal} {Phys.Rev.}\ }\textbf {\bibinfo {volume} {D88}},\ \bibinfo {pages}
  {044047} (\bibinfo {year} {2013}{\natexlab{a}})},\ \Eprint
  {http://arxiv.org/abs/1307.8086} {arXiv:1307.8086 [gr-qc]} \BibitemShut
  {NoStop}%
\bibitem [{\citenamefont {Gralla}\ \emph
  {et~al.}(2016{\natexlab{a}})\citenamefont {Gralla}, \citenamefont {Hughes},\
  and\ \citenamefont {Warburton}}]{Gralla:2016qfw}%
  \BibitemOpen
  \bibfield  {author} {\bibinfo {author} {\bibfnamefont {S.~E.}\ \bibnamefont
  {Gralla}}, \bibinfo {author} {\bibfnamefont {S.~A.}\ \bibnamefont {Hughes}},
  \ and\ \bibinfo {author} {\bibfnamefont {N.}~\bibnamefont {Warburton}},\
  }\href {\doibase 10.1088/0264-9381/33/15/155002} {\bibfield  {journal}
  {\bibinfo  {journal} {Class. Quant. Grav.}\ }\textbf {\bibinfo {volume}
  {33}},\ \bibinfo {pages} {155002} (\bibinfo {year} {2016}{\natexlab{a}})},\
  \Eprint {http://arxiv.org/abs/1603.01221} {arXiv:1603.01221 [gr-qc]}
  \BibitemShut {NoStop}%
\bibitem [{\citenamefont {Burko}\ and\ \citenamefont
  {Khanna}(2016)}]{Burko:2016sfi}%
  \BibitemOpen
  \bibfield  {author} {\bibinfo {author} {\bibfnamefont {L.~M.}\ \bibnamefont
  {Burko}}\ and\ \bibinfo {author} {\bibfnamefont {G.}~\bibnamefont {Khanna}},\
  }\href@noop {} {\  (\bibinfo {year} {2016})},\ \Eprint
  {http://arxiv.org/abs/1608.02244} {arXiv:1608.02244 [gr-qc]} \BibitemShut
  {NoStop}%
\bibitem [{\citenamefont {{Guica}}\ \emph {et~al.}(2009)\citenamefont
  {{Guica}}, \citenamefont {{Hartman}}, \citenamefont {{Song}},\ and\
  \citenamefont {{Strominger}}}]{kerrCFT}%
  \BibitemOpen
  \bibfield  {author} {\bibinfo {author} {\bibfnamefont {M.}~\bibnamefont
  {{Guica}}}, \bibinfo {author} {\bibfnamefont {T.}~\bibnamefont {{Hartman}}},
  \bibinfo {author} {\bibfnamefont {W.}~\bibnamefont {{Song}}}, \ and\ \bibinfo
  {author} {\bibfnamefont {A.}~\bibnamefont {{Strominger}}},\ }\href {\doibase
  10.1103/PhysRevD.80.124008} {\bibfield  {journal} {\bibinfo  {journal}
  {\prd}\ }\textbf {\bibinfo {volume} {80}},\ \bibinfo {eid} {124008} (\bibinfo
  {year} {2009})},\ \Eprint {http://arxiv.org/abs/0809.4266} {arXiv:0809.4266
  [hep-th]} \BibitemShut {NoStop}%
\bibitem [{\citenamefont {Aretakis}(2010)}]{Aretakis:2010gd}%
  \BibitemOpen
  \bibfield  {author} {\bibinfo {author} {\bibfnamefont {S.}~\bibnamefont
  {Aretakis}},\ }\href@noop {} {\  (\bibinfo {year} {2010})},\ \Eprint
  {http://arxiv.org/abs/1006.0283} {arXiv:1006.0283 [math.AP]} \BibitemShut
  {NoStop}%
\bibitem [{\citenamefont {Aretakis}(2011{\natexlab{a}})}]{Aretakis:2011ha}%
  \BibitemOpen
  \bibfield  {author} {\bibinfo {author} {\bibfnamefont {S.}~\bibnamefont
  {Aretakis}},\ }\href {\doibase 10.1007/s00220-011-1254-5} {\bibfield
  {journal} {\bibinfo  {journal} {Commun. Math. Phys.}\ }\textbf {\bibinfo
  {volume} {307}},\ \bibinfo {pages} {17} (\bibinfo {year}
  {2011}{\natexlab{a}})},\ \Eprint {http://arxiv.org/abs/1110.2007}
  {arXiv:1110.2007 [gr-qc]} \BibitemShut {NoStop}%
\bibitem [{\citenamefont {Aretakis}(2011{\natexlab{b}})}]{Aretakis:2011hc}%
  \BibitemOpen
  \bibfield  {author} {\bibinfo {author} {\bibfnamefont {S.}~\bibnamefont
  {Aretakis}},\ }\href {\doibase 10.1007/s00023-011-0110-7} {\bibfield
  {journal} {\bibinfo  {journal} {Annales Henri Poincare}\ }\textbf {\bibinfo
  {volume} {12}},\ \bibinfo {pages} {1491} (\bibinfo {year}
  {2011}{\natexlab{b}})},\ \Eprint {http://arxiv.org/abs/1110.2009}
  {arXiv:1110.2009 [gr-qc]} \BibitemShut {NoStop}%
\bibitem [{\citenamefont {Dafermos}\ and\ \citenamefont
  {Rodnianski}(2009)}]{Dafermos:2005eh}%
  \BibitemOpen
  \bibfield  {author} {\bibinfo {author} {\bibfnamefont {M.}~\bibnamefont
  {Dafermos}}\ and\ \bibinfo {author} {\bibfnamefont {I.}~\bibnamefont
  {Rodnianski}},\ }\href@noop {} {\bibfield  {journal} {\bibinfo  {journal}
  {Commun. Pure Appl. Math.}\ }\textbf {\bibinfo {volume} {62}},\ \bibinfo
  {pages} {859} (\bibinfo {year} {2009})},\ \Eprint
  {http://arxiv.org/abs/gr-qc/0512119} {arXiv:gr-qc/0512119 [gr-qc]}
  \BibitemShut {NoStop}%
\bibitem [{\citenamefont {Aretakis}(2013)}]{Aretakis:2013oja}%
  \BibitemOpen
  \bibfield  {author} {\bibinfo {author} {\bibfnamefont {S.}~\bibnamefont
  {Aretakis}},\ }\href@noop {} {\  (\bibinfo {year} {2013})},\ \Eprint
  {http://arxiv.org/abs/1310.1365} {arXiv:1310.1365 [gr-qc]} \BibitemShut
  {NoStop}%
\bibitem [{\citenamefont {{Lucietti}}\ and\ \citenamefont
  {{Reall}}(2012)}]{Lucietti2012}%
  \BibitemOpen
  \bibfield  {author} {\bibinfo {author} {\bibfnamefont {J.}~\bibnamefont
  {{Lucietti}}}\ and\ \bibinfo {author} {\bibfnamefont {H.~S.}\ \bibnamefont
  {{Reall}}},\ }\href {\doibase 10.1103/PhysRevD.86.104030} {\bibfield
  {journal} {\bibinfo  {journal} {\prd}\ }\textbf {\bibinfo {volume} {86}},\
  \bibinfo {eid} {104030} (\bibinfo {year} {2012})},\ \Eprint
  {http://arxiv.org/abs/1208.1437} {arXiv:1208.1437 [gr-qc]} \BibitemShut
  {NoStop}%
\bibitem [{\citenamefont {Aretakis}(2015)}]{Aretakis:2012ei}%
  \BibitemOpen
  \bibfield  {author} {\bibinfo {author} {\bibfnamefont {S.}~\bibnamefont
  {Aretakis}},\ }\href {\doibase 10.4310/ATMP.2015.v19.n3.a1} {\bibfield
  {journal} {\bibinfo  {journal} {Adv. Theor. Math. Phys.}\ }\textbf {\bibinfo
  {volume} {19}},\ \bibinfo {pages} {507} (\bibinfo {year} {2015})},\ \Eprint
  {http://arxiv.org/abs/1206.6598} {arXiv:1206.6598 [gr-qc]} \BibitemShut
  {NoStop}%
\bibitem [{\citenamefont {Murata}(2013)}]{Murata:2012ct}%
  \BibitemOpen
  \bibfield  {author} {\bibinfo {author} {\bibfnamefont {K.}~\bibnamefont
  {Murata}},\ }\href {\doibase 10.1088/0264-9381/30/7/075002} {\bibfield
  {journal} {\bibinfo  {journal} {Class. Quant. Grav.}\ }\textbf {\bibinfo
  {volume} {30}},\ \bibinfo {pages} {075002} (\bibinfo {year} {2013})},\
  \Eprint {http://arxiv.org/abs/1211.6903} {arXiv:1211.6903 [gr-qc]}
  \BibitemShut {NoStop}%
\bibitem [{\citenamefont {Casals}\ \emph {et~al.}(2016)\citenamefont {Casals},
  \citenamefont {Gralla},\ and\ \citenamefont {Zimmerman}}]{Casals:2016mel}%
  \BibitemOpen
  \bibfield  {author} {\bibinfo {author} {\bibfnamefont {M.}~\bibnamefont
  {Casals}}, \bibinfo {author} {\bibfnamefont {S.~E.}\ \bibnamefont {Gralla}},
  \ and\ \bibinfo {author} {\bibfnamefont {P.}~\bibnamefont {Zimmerman}},\
  }\href {\doibase 10.1103/PhysRevD.94.064003} {\bibfield  {journal} {\bibinfo
  {journal} {Phys. Rev.}\ }\textbf {\bibinfo {volume} {D94}},\ \bibinfo {pages}
  {064003} (\bibinfo {year} {2016})},\ \Eprint
  {http://arxiv.org/abs/1606.08505} {arXiv:1606.08505 [gr-qc]} \BibitemShut
  {NoStop}%
\bibitem [{\citenamefont {Israel}(1986)}]{Israel:1986gqz}%
  \BibitemOpen
  \bibfield  {author} {\bibinfo {author} {\bibfnamefont {W.}~\bibnamefont
  {Israel}},\ }\href {\doibase 10.1103/PhysRevLett.57.397} {\bibfield
  {journal} {\bibinfo  {journal} {Phys. Rev. Lett.}\ }\textbf {\bibinfo
  {volume} {57}},\ \bibinfo {pages} {397} (\bibinfo {year} {1986})}\BibitemShut
  {NoStop}%
\bibitem [{\citenamefont {Murata}\ \emph {et~al.}(2013)\citenamefont {Murata},
  \citenamefont {Reall},\ and\ \citenamefont {Tanahashi}}]{Murata:2013daa}%
  \BibitemOpen
  \bibfield  {author} {\bibinfo {author} {\bibfnamefont {K.}~\bibnamefont
  {Murata}}, \bibinfo {author} {\bibfnamefont {H.~S.}\ \bibnamefont {Reall}}, \
  and\ \bibinfo {author} {\bibfnamefont {N.}~\bibnamefont {Tanahashi}},\ }\href
  {\doibase 10.1088/0264-9381/30/23/235007} {\bibfield  {journal} {\bibinfo
  {journal} {Class. Quant. Grav.}\ }\textbf {\bibinfo {volume} {30}},\ \bibinfo
  {pages} {235007} (\bibinfo {year} {2013})},\ \Eprint
  {http://arxiv.org/abs/1307.6800} {arXiv:1307.6800 [gr-qc]} \BibitemShut
  {NoStop}%
\bibitem [{\citenamefont {Gralla}\ \emph
  {et~al.}(2016{\natexlab{b}})\citenamefont {Gralla}, \citenamefont
  {Zimmerman},\ and\ \citenamefont {Zimmerman}}]{Gralla:2016sxp}%
  \BibitemOpen
  \bibfield  {author} {\bibinfo {author} {\bibfnamefont {S.~E.}\ \bibnamefont
  {Gralla}}, \bibinfo {author} {\bibfnamefont {A.}~\bibnamefont {Zimmerman}}, \
  and\ \bibinfo {author} {\bibfnamefont {P.}~\bibnamefont {Zimmerman}},\ }\href
  {\doibase 10.1103/PhysRevD.94.084017} {\bibfield  {journal} {\bibinfo
  {journal} {Phys. Rev.}\ }\textbf {\bibinfo {volume} {D94}},\ \bibinfo {pages}
  {084017} (\bibinfo {year} {2016}{\natexlab{b}})},\ \Eprint
  {http://arxiv.org/abs/1608.04739} {arXiv:1608.04739 [gr-qc]} \BibitemShut
  {NoStop}%
\bibitem [{\citenamefont {Baier}\ \emph {et~al.}(2008)\citenamefont {Baier},
  \citenamefont {Romatschke}, \citenamefont {Son}, \citenamefont {Starinets},\
  and\ \citenamefont {Stephanov}}]{Baier:2007ix}%
  \BibitemOpen
  \bibfield  {author} {\bibinfo {author} {\bibfnamefont {R.}~\bibnamefont
  {Baier}}, \bibinfo {author} {\bibfnamefont {P.}~\bibnamefont {Romatschke}},
  \bibinfo {author} {\bibfnamefont {D.~T.}\ \bibnamefont {Son}}, \bibinfo
  {author} {\bibfnamefont {A.~O.}\ \bibnamefont {Starinets}}, \ and\ \bibinfo
  {author} {\bibfnamefont {M.~A.}\ \bibnamefont {Stephanov}},\ }\href {\doibase
  10.1088/1126-6708/2008/04/100} {\bibfield  {journal} {\bibinfo  {journal}
  {JHEP}\ }\textbf {\bibinfo {volume} {04}},\ \bibinfo {pages} {100} (\bibinfo
  {year} {2008})},\ \Eprint {http://arxiv.org/abs/0712.2451} {arXiv:0712.2451
  [hep-th]} \BibitemShut {NoStop}%
\bibitem [{\citenamefont {Bhattacharyya}\ \emph {et~al.}(2008)\citenamefont
  {Bhattacharyya}, \citenamefont {Hubeny}, \citenamefont {Minwalla},\ and\
  \citenamefont {Rangamani}}]{Bhattacharyya:2008jc}%
  \BibitemOpen
  \bibfield  {author} {\bibinfo {author} {\bibfnamefont {S.}~\bibnamefont
  {Bhattacharyya}}, \bibinfo {author} {\bibfnamefont {V.~E.}\ \bibnamefont
  {Hubeny}}, \bibinfo {author} {\bibfnamefont {S.}~\bibnamefont {Minwalla}}, \
  and\ \bibinfo {author} {\bibfnamefont {M.}~\bibnamefont {Rangamani}},\ }\href
  {\doibase 10.1088/1126-6708/2008/02/045} {\bibfield  {journal} {\bibinfo
  {journal} {JHEP}\ }\textbf {\bibinfo {volume} {02}},\ \bibinfo {pages} {045}
  (\bibinfo {year} {2008})},\ \Eprint {http://arxiv.org/abs/0712.2456}
  {arXiv:0712.2456 [hep-th]} \BibitemShut {NoStop}%
\bibitem [{\citenamefont {Adams}\ \emph {et~al.}(2014)\citenamefont {Adams},
  \citenamefont {Chesler},\ and\ \citenamefont {Liu}}]{Adams:2013vsa}%
  \BibitemOpen
  \bibfield  {author} {\bibinfo {author} {\bibfnamefont {A.}~\bibnamefont
  {Adams}}, \bibinfo {author} {\bibfnamefont {P.~M.}\ \bibnamefont {Chesler}},
  \ and\ \bibinfo {author} {\bibfnamefont {H.}~\bibnamefont {Liu}},\ }\href
  {\doibase 10.1103/PhysRevLett.112.151602} {\bibfield  {journal} {\bibinfo
  {journal} {Phys. Rev. Lett.}\ }\textbf {\bibinfo {volume} {112}},\ \bibinfo
  {pages} {151602} (\bibinfo {year} {2014})},\ \Eprint
  {http://arxiv.org/abs/1307.7267} {arXiv:1307.7267 [hep-th]} \BibitemShut
  {NoStop}%
\bibitem [{\citenamefont {Green}\ \emph {et~al.}(2014)\citenamefont {Green},
  \citenamefont {Carrasco},\ and\ \citenamefont {Lehner}}]{Green:2013zba}%
  \BibitemOpen
  \bibfield  {author} {\bibinfo {author} {\bibfnamefont {S.~R.}\ \bibnamefont
  {Green}}, \bibinfo {author} {\bibfnamefont {F.}~\bibnamefont {Carrasco}}, \
  and\ \bibinfo {author} {\bibfnamefont {L.}~\bibnamefont {Lehner}},\ }\href
  {\doibase 10.1103/PhysRevX.4.011001} {\bibfield  {journal} {\bibinfo
  {journal} {Phys. Rev.}\ }\textbf {\bibinfo {volume} {X4}},\ \bibinfo {pages}
  {011001} (\bibinfo {year} {2014})},\ \Eprint {http://arxiv.org/abs/1309.7940}
  {arXiv:1309.7940 [hep-th]} \BibitemShut {NoStop}%
\bibitem [{\citenamefont {Hartnoll}(2009)}]{Hartnoll:2009sz}%
  \BibitemOpen
  \bibfield  {author} {\bibinfo {author} {\bibfnamefont {S.~A.}\ \bibnamefont
  {Hartnoll}},\ }\bibfield  {booktitle} {\emph {\bibinfo {booktitle} {{Strings,
  Supergravity and Gauge Theories. Proceedings, CERN Winter School, CERN,
  Geneva, Switzerland, February 9-13 2009}}},\ }\href {\doibase
  10.1088/0264-9381/26/22/224002} {\bibfield  {journal} {\bibinfo  {journal}
  {Class. Quant. Grav.}\ }\textbf {\bibinfo {volume} {26}},\ \bibinfo {pages}
  {224002} (\bibinfo {year} {2009})},\ \Eprint {http://arxiv.org/abs/0903.3246}
  {arXiv:0903.3246 [hep-th]} \BibitemShut {NoStop}%
\bibitem [{\citenamefont {Gubser}(2008)}]{Gubser:2008px}%
  \BibitemOpen
  \bibfield  {author} {\bibinfo {author} {\bibfnamefont {S.~S.}\ \bibnamefont
  {Gubser}},\ }\href {\doibase 10.1103/PhysRevD.78.065034} {\bibfield
  {journal} {\bibinfo  {journal} {Phys. Rev.}\ }\textbf {\bibinfo {volume}
  {D78}},\ \bibinfo {pages} {065034} (\bibinfo {year} {2008})},\ \Eprint
  {http://arxiv.org/abs/0801.2977} {arXiv:0801.2977 [hep-th]} \BibitemShut
  {NoStop}%
\bibitem [{\citenamefont {Hartnoll}\ \emph {et~al.}(2008)\citenamefont
  {Hartnoll}, \citenamefont {Herzog},\ and\ \citenamefont
  {Horowitz}}]{Hartnoll:2008vx}%
  \BibitemOpen
  \bibfield  {author} {\bibinfo {author} {\bibfnamefont {S.~A.}\ \bibnamefont
  {Hartnoll}}, \bibinfo {author} {\bibfnamefont {C.~P.}\ \bibnamefont
  {Herzog}}, \ and\ \bibinfo {author} {\bibfnamefont {G.~T.}\ \bibnamefont
  {Horowitz}},\ }\href {\doibase 10.1103/PhysRevLett.101.031601} {\bibfield
  {journal} {\bibinfo  {journal} {Phys. Rev. Lett.}\ }\textbf {\bibinfo
  {volume} {101}},\ \bibinfo {pages} {031601} (\bibinfo {year} {2008})},\
  \Eprint {http://arxiv.org/abs/0803.3295} {arXiv:0803.3295 [hep-th]}
  \BibitemShut {NoStop}%
\bibitem [{\citenamefont {Barut}\ and\ \citenamefont
  {Fronsdal}(1965)}]{Barut:1965}%
  \BibitemOpen
  \bibfield  {author} {\bibinfo {author} {\bibfnamefont {A.~O.}\ \bibnamefont
  {Barut}}\ and\ \bibinfo {author} {\bibfnamefont {C.}~\bibnamefont
  {Fronsdal}},\ }\href@noop {} {\bibfield  {journal} {\bibinfo  {journal}
  {Proceedings of the Royal Society of London. Series A, Mathematical and
  Physical Sciences}\ }\textbf {\bibinfo {volume} {287}},\ \bibinfo {pages}
  {532} (\bibinfo {year} {1965})}\BibitemShut {NoStop}%
\bibitem [{\citenamefont {{Detweiler}}(1977)}]{Detweiler1977}%
  \BibitemOpen
  \bibfield  {author} {\bibinfo {author} {\bibfnamefont {S.}~\bibnamefont
  {{Detweiler}}},\ }\href {\doibase 10.1098/rspa.1977.0005} {\bibfield
  {journal} {\bibinfo  {journal} {Royal Society of London Proceedings Series
  A}\ }\textbf {\bibinfo {volume} {352}},\ \bibinfo {pages} {381} (\bibinfo
  {year} {1977})}\BibitemShut {NoStop}%
\bibitem [{\citenamefont {Leaver}(1985)}]{Leaver1985}%
  \BibitemOpen
  \bibfield  {author} {\bibinfo {author} {\bibfnamefont {E.}~\bibnamefont
  {Leaver}},\ }\href@noop {} {\bibfield  {journal} {\bibinfo  {journal} {Proc.
  Roy. Soc. Lond. A}\ }\textbf {\bibinfo {volume} {402}},\ \bibinfo {pages}
  {285} (\bibinfo {year} {1985})}\BibitemShut {NoStop}%
\bibitem [{\citenamefont {{Cardoso}}(2004)}]{Cardoso2004}%
  \BibitemOpen
  \bibfield  {author} {\bibinfo {author} {\bibfnamefont {V.}~\bibnamefont
  {{Cardoso}}},\ }\href {\doibase 10.1103/PhysRevD.70.127502} {\bibfield
  {journal} {\bibinfo  {journal} {\prd}\ }\textbf {\bibinfo {volume} {70}},\
  \bibinfo {eid} {127502} (\bibinfo {year} {2004})},\ \Eprint
  {http://arxiv.org/abs/arXiv:gr-qc/0411048} {arXiv:gr-qc/0411048} \BibitemShut
  {NoStop}%
\bibitem [{\citenamefont {Hod}(2010)}]{Hod2010chargedfields}%
  \BibitemOpen
  \bibfield  {author} {\bibinfo {author} {\bibfnamefont {S.}~\bibnamefont
  {Hod}},\ }\href {\doibase http://dx.doi.org/10.1016/j.physleta.2010.05.052}
  {\bibfield  {journal} {\bibinfo  {journal} {Physics Letters A}\ }\textbf
  {\bibinfo {volume} {374}},\ \bibinfo {pages} {2901 } (\bibinfo {year}
  {2010})}\BibitemShut {NoStop}%
\bibitem [{\citenamefont {Hod}(2012)}]{Hod2012349}%
  \BibitemOpen
  \bibfield  {author} {\bibinfo {author} {\bibfnamefont {S.}~\bibnamefont
  {Hod}},\ }\href {\doibase http://dx.doi.org/10.1016/j.physletb.2012.03.010}
  {\bibfield  {journal} {\bibinfo  {journal} {Physics Letters B}\ }\textbf
  {\bibinfo {volume} {710}},\ \bibinfo {pages} {349 } (\bibinfo {year}
  {2012})}\BibitemShut {NoStop}%
\bibitem [{\citenamefont {{Hod}}(2012)}]{HodEikonal2012}%
  \BibitemOpen
  \bibfield  {author} {\bibinfo {author} {\bibfnamefont {S.}~\bibnamefont
  {{Hod}}},\ }\href {\doibase 10.1016/j.physletb.2012.08.001} {\bibfield
  {journal} {\bibinfo  {journal} {Physics Letters B}\ }\textbf {\bibinfo
  {volume} {715}},\ \bibinfo {pages} {348} (\bibinfo {year} {2012})},\ \Eprint
  {http://arxiv.org/abs/1207.5282} {arXiv:1207.5282 [gr-qc]} \BibitemShut
  {NoStop}%
\bibitem [{\citenamefont {Cook}\ and\ \citenamefont
  {Zalutskiy}(2014)}]{Cook:2014cta}%
  \BibitemOpen
  \bibfield  {author} {\bibinfo {author} {\bibfnamefont {G.~B.}\ \bibnamefont
  {Cook}}\ and\ \bibinfo {author} {\bibfnamefont {M.}~\bibnamefont
  {Zalutskiy}},\ }\href {\doibase 10.1103/PhysRevD.90.124021} {\bibfield
  {journal} {\bibinfo  {journal} {Phys. Rev.}\ }\textbf {\bibinfo {volume}
  {D90}},\ \bibinfo {pages} {124021} (\bibinfo {year} {2014})},\ \Eprint
  {http://arxiv.org/abs/1410.7698} {arXiv:1410.7698 [gr-qc]} \BibitemShut
  {NoStop}%
\bibitem [{\citenamefont {Richartz}\ and\ \citenamefont
  {Giugno}(2014)}]{Richartz:2014jla}%
  \BibitemOpen
  \bibfield  {author} {\bibinfo {author} {\bibfnamefont {M.}~\bibnamefont
  {Richartz}}\ and\ \bibinfo {author} {\bibfnamefont {D.}~\bibnamefont
  {Giugno}},\ }\href {\doibase 10.1103/PhysRevD.90.124011} {\bibfield
  {journal} {\bibinfo  {journal} {Phys. Rev.}\ }\textbf {\bibinfo {volume}
  {D90}},\ \bibinfo {pages} {124011} (\bibinfo {year} {2014})},\ \Eprint
  {http://arxiv.org/abs/1409.7440} {arXiv:1409.7440 [gr-qc]} \BibitemShut
  {NoStop}%
\bibitem [{\citenamefont {Konoplya}\ and\ \citenamefont
  {Zhidenko}(2013)}]{PhysRevD.88.024054}%
  \BibitemOpen
  \bibfield  {author} {\bibinfo {author} {\bibfnamefont {R.~A.}\ \bibnamefont
  {Konoplya}}\ and\ \bibinfo {author} {\bibfnamefont {A.}~\bibnamefont
  {Zhidenko}},\ }\href {\doibase 10.1103/PhysRevD.88.024054} {\bibfield
  {journal} {\bibinfo  {journal} {Phys. Rev. D}\ }\textbf {\bibinfo {volume}
  {88}},\ \bibinfo {pages} {024054} (\bibinfo {year} {2013})}\BibitemShut
  {NoStop}%
\bibitem [{\citenamefont {Richartz}(2016)}]{Richartz:2016}%
  \BibitemOpen
  \bibfield  {author} {\bibinfo {author} {\bibfnamefont {M.}~\bibnamefont
  {Richartz}},\ }\href {\doibase 10.1103/PhysRevD.93.064062} {\bibfield
  {journal} {\bibinfo  {journal} {Phys. Rev. D}\ }\textbf {\bibinfo {volume}
  {93}},\ \bibinfo {pages} {064062} (\bibinfo {year} {2016})}\BibitemShut
  {NoStop}%
\bibitem [{\citenamefont {Misner}\ \emph {et~al.}(1973)\citenamefont {Misner},
  \citenamefont {Thorne},\ and\ \citenamefont {Wheeler}}]{MTW}%
  \BibitemOpen
  \bibfield  {author} {\bibinfo {author} {\bibfnamefont {C.}~\bibnamefont
  {Misner}}, \bibinfo {author} {\bibfnamefont {K.}~\bibnamefont {Thorne}}, \
  and\ \bibinfo {author} {\bibfnamefont {J.}~\bibnamefont {Wheeler}},\
  }\href@noop {} {\emph {\bibinfo {title} {Gravitation}}}\ (\bibinfo
  {publisher} {{Freeman}},\ \bibinfo {address} {{San Francisco}},\ \bibinfo
  {year} {1973})\BibitemShut {NoStop}%
\bibitem [{\citenamefont {Robinson}(1959)}]{robinson1959solution}%
  \BibitemOpen
  \bibfield  {author} {\bibinfo {author} {\bibfnamefont {I.}~\bibnamefont
  {Robinson}},\ }\href@noop {} {\bibfield  {journal} {\bibinfo  {journal}
  {Bull. Acad. Pol. Sci. Ser. Sci. Math. Astron. Phys}\ }\textbf {\bibinfo
  {volume} {7}},\ \bibinfo {pages} {351} (\bibinfo {year} {1959})}\BibitemShut
  {NoStop}%
\bibitem [{\citenamefont {Bertotti}(1959)}]{Bertotti1959}%
  \BibitemOpen
  \bibfield  {author} {\bibinfo {author} {\bibfnamefont {B.}~\bibnamefont
  {Bertotti}},\ }\href {\doibase 10.1103/PhysRev.116.1331} {\bibfield
  {journal} {\bibinfo  {journal} {Phys. Rev.}\ }\textbf {\bibinfo {volume}
  {116}},\ \bibinfo {pages} {1331} (\bibinfo {year} {1959})}\BibitemShut
  {NoStop}%
\bibitem [{\citenamefont {Maldacena}\ \emph {et~al.}(1999)\citenamefont
  {Maldacena}, \citenamefont {Michelson},\ and\ \citenamefont
  {Strominger}}]{Maldacena:1998uz}%
  \BibitemOpen
  \bibfield  {author} {\bibinfo {author} {\bibfnamefont {J.~M.}\ \bibnamefont
  {Maldacena}}, \bibinfo {author} {\bibfnamefont {J.}~\bibnamefont
  {Michelson}}, \ and\ \bibinfo {author} {\bibfnamefont {A.}~\bibnamefont
  {Strominger}},\ }\href {\doibase 10.1088/1126-6708/1999/02/011} {\bibfield
  {journal} {\bibinfo  {journal} {JHEP}\ }\textbf {\bibinfo {volume} {02}},\
  \bibinfo {pages} {011} (\bibinfo {year} {1999})},\ \Eprint
  {http://arxiv.org/abs/hep-th/9812073} {arXiv:hep-th/9812073 [hep-th]}
  \BibitemShut {NoStop}%
\bibitem [{\citenamefont {Carter}(2009)}]{Carter2009}%
  \BibitemOpen
  \bibfield  {author} {\bibinfo {author} {\bibfnamefont {B.}~\bibnamefont
  {Carter}},\ }\href {\doibase 10.1007/s10714-009-0888-5} {\bibfield  {journal}
  {\bibinfo  {journal} {General Relativity and Gravitation}\ }\textbf {\bibinfo
  {volume} {41}},\ \bibinfo {pages} {2873} (\bibinfo {year}
  {2009})}\BibitemShut {NoStop}%
\bibitem [{\citenamefont {Olver}\ \emph {et~al.}(2010)\citenamefont {Olver},
  \citenamefont {Lozier}, \citenamefont {Boisvert},\ and\ \citenamefont
  {Clark}}]{nist}%
  \BibitemOpen
  \bibfield  {author} {\bibinfo {author} {\bibfnamefont {F.~W.}\ \bibnamefont
  {Olver}}, \bibinfo {author} {\bibfnamefont {D.~W.}\ \bibnamefont {Lozier}},
  \bibinfo {author} {\bibfnamefont {R.~F.}\ \bibnamefont {Boisvert}}, \ and\
  \bibinfo {author} {\bibfnamefont {C.~W.}\ \bibnamefont {Clark}},\ }\href
  {http://dlmf.nist.gov/} {\emph {\bibinfo {title} {NIST Handbook of
  Mathematical Functions}}},\ \bibinfo {edition} {1st}\ ed.\ (\bibinfo
  {publisher} {Cambridge University Press},\ \bibinfo {address} {New York, NY,
  USA},\ \bibinfo {year} {2010})\BibitemShut {NoStop}%
\bibitem [{\citenamefont {Spradlin}\ and\ \citenamefont
  {Strominger}(1999)}]{Spradlin:1999bn}%
  \BibitemOpen
  \bibfield  {author} {\bibinfo {author} {\bibfnamefont {M.}~\bibnamefont
  {Spradlin}}\ and\ \bibinfo {author} {\bibfnamefont {A.}~\bibnamefont
  {Strominger}},\ }\href {\doibase 10.1088/1126-6708/1999/11/021} {\bibfield
  {journal} {\bibinfo  {journal} {JHEP}\ }\textbf {\bibinfo {volume} {11}},\
  \bibinfo {pages} {021} (\bibinfo {year} {1999})},\ \Eprint
  {http://arxiv.org/abs/hep-th/9904143} {arXiv:hep-th/9904143 [hep-th]}
  \BibitemShut {NoStop}%
\bibitem [{\citenamefont {{Leaver}}(1986)}]{Leaver1986b}%
  \BibitemOpen
  \bibfield  {author} {\bibinfo {author} {\bibfnamefont {E.~W.}\ \bibnamefont
  {{Leaver}}},\ }\href {\doibase 10.1103/PhysRevD.34.384} {\bibfield  {journal}
  {\bibinfo  {journal} {\prd}\ }\textbf {\bibinfo {volume} {34}},\ \bibinfo
  {pages} {384} (\bibinfo {year} {1986})}\BibitemShut {NoStop}%
\bibitem [{\citenamefont {{Hod}}(2009)}]{Hod2009}%
  \BibitemOpen
  \bibfield  {author} {\bibinfo {author} {\bibfnamefont {S.}~\bibnamefont
  {{Hod}}},\ }\href {\doibase 10.1103/PhysRevD.80.064004} {\bibfield  {journal}
  {\bibinfo  {journal} {\prd}\ }\textbf {\bibinfo {volume} {80}},\ \bibinfo
  {eid} {064004} (\bibinfo {year} {2009})},\ \Eprint
  {http://arxiv.org/abs/0909.0314} {arXiv:0909.0314 [gr-qc]} \BibitemShut
  {NoStop}%
\bibitem [{\citenamefont {Yang}\ \emph
  {et~al.}(2013{\natexlab{b}})\citenamefont {Yang}, \citenamefont {Zhang},
  \citenamefont {Zimmerman}, \citenamefont {Nichols}, \citenamefont {Berti}
  \emph {et~al.}}]{Yang:2012pj}%
  \BibitemOpen
  \bibfield  {author} {\bibinfo {author} {\bibfnamefont {H.}~\bibnamefont
  {Yang}}, \bibinfo {author} {\bibfnamefont {F.}~\bibnamefont {Zhang}},
  \bibinfo {author} {\bibfnamefont {A.}~\bibnamefont {Zimmerman}}, \bibinfo
  {author} {\bibfnamefont {D.~A.}\ \bibnamefont {Nichols}}, \bibinfo {author}
  {\bibfnamefont {E.}~\bibnamefont {Berti}},  \emph {et~al.},\ }\href {\doibase
  10.1103/PhysRevD.87.041502} {\bibfield  {journal} {\bibinfo  {journal}
  {Phys.Rev.}\ }\textbf {\bibinfo {volume} {D87}},\ \bibinfo {pages} {041502}
  (\bibinfo {year} {2013}{\natexlab{b}})},\ \Eprint
  {http://arxiv.org/abs/1212.3271} {arXiv:1212.3271 [gr-qc]} \BibitemShut
  {NoStop}%
\bibitem [{\citenamefont {Doetsch}(1974)}]{Doetsch1974}%
  \BibitemOpen
  \bibfield  {author} {\bibinfo {author} {\bibfnamefont {G.}~\bibnamefont
  {Doetsch}},\ }\href {\doibase 10.1007/978-3-642-65690-3_36} {\emph {\bibinfo
  {title} {Introduction to the Theory and Application of the Laplace
  Transformation}}}\ (\bibinfo  {publisher} {Springer Berlin Heidelberg},\
  \bibinfo {address} {Berlin, Heidelberg},\ \bibinfo {year} {1974})\BibitemShut
  {NoStop}%
\bibitem [{\citenamefont {Casals}\ \emph {et~al.}()\citenamefont {Casals},
  \citenamefont {Gralla},\ and\ \citenamefont {Zimmerman}}]{CGZsoon}%
  \BibitemOpen
  \bibfield  {author} {\bibinfo {author} {\bibfnamefont {M.}~\bibnamefont
  {Casals}}, \bibinfo {author} {\bibfnamefont {S.~E.}\ \bibnamefont {Gralla}},
  \ and\ \bibinfo {author} {\bibfnamefont {P.}~\bibnamefont {Zimmerman}},\
  }\href@noop {} {\bibinfo  {journal} {in preparation}\ }\BibitemShut {NoStop}%
\bibitem [{\citenamefont {Valk{\'o}}\ and\ \citenamefont
  {Abate}(2004)}]{VALKO2004629}%
  \BibitemOpen
\bibfield  {journal} {  }\bibfield  {author} {\bibinfo {author} {\bibfnamefont
  {P.}~\bibnamefont {Valk{\'o}}}\ and\ \bibinfo {author} {\bibfnamefont
  {J.}~\bibnamefont {Abate}},\ }\href
  {http://www.sciencedirect.com/science/article/pii/S0898122104840823}
  {\bibfield  {journal} {\bibinfo  {journal} {Computers and Mathematics with
  Applications}\ }\textbf {\bibinfo {volume} {48}},\ \bibinfo {pages} {629 }
  (\bibinfo {year} {2004})}\BibitemShut {NoStop}%
\bibitem [{\citenamefont {Prabhu}(2015)}]{Prabhu:2015vua}%
  \BibitemOpen
  \bibfield  {author} {\bibinfo {author} {\bibfnamefont {K.}~\bibnamefont
  {Prabhu}},\ }\href@noop {} {\  (\bibinfo {year} {2015})},\ \Eprint
  {http://arxiv.org/abs/1511.00388} {arXiv:1511.00388 [gr-qc]} \BibitemShut
  {NoStop}%
\bibitem [{\citenamefont {Wybourne}(1974)}]{Wybourne1974book}%
  \BibitemOpen
  \bibfield  {author} {\bibinfo {author} {\bibfnamefont {B.~G.}\ \bibnamefont
  {Wybourne}},\ }\href@noop {} {\emph {\bibinfo {title} {Classical Groups for
  Physicists, by Group Theory for Physicists}}},\ Wiley-Interscience\ (\bibinfo
   {publisher} {John Wiley and Sons},\ \bibinfo {year} {1974})\BibitemShut
  {NoStop}%
\bibitem [{\citenamefont {Bargmann}(1947)}]{Bargmann1947}%
  \BibitemOpen
  \bibfield  {author} {\bibinfo {author} {\bibfnamefont {V.}~\bibnamefont
  {Bargmann}},\ }\href {http://www.jstor.org/stable/1969129} {\bibfield
  {journal} {\bibinfo  {journal} {Annals of Mathematics}\ }\textbf {\bibinfo
  {volume} {48}},\ \bibinfo {pages} {568} (\bibinfo {year} {1947})}\BibitemShut
  {NoStop}%
\bibitem [{\citenamefont {Gitman}\ and\ \citenamefont
  {Shelepin}(1997)}]{Gitman1997}%
  \BibitemOpen
  \bibfield  {author} {\bibinfo {author} {\bibfnamefont {D.~M.}\ \bibnamefont
  {Gitman}}\ and\ \bibinfo {author} {\bibfnamefont {A.~L.}\ \bibnamefont
  {Shelepin}},\ }\href {http://stacks.iop.org/0305-4470/30/i=17/a=018}
  {\bibfield  {journal} {\bibinfo  {journal} {Journal of Physics A:
  Mathematical and General}\ }\textbf {\bibinfo {volume} {30}},\ \bibinfo
  {pages} {6093} (\bibinfo {year} {1997})}\BibitemShut {NoStop}%
\bibitem [{\citenamefont {Balasubramanian}\ \emph {et~al.}(1999)\citenamefont
  {Balasubramanian}, \citenamefont {Kraus},\ and\ \citenamefont
  {Lawrence}}]{Balasubramanian:1998sn}%
  \BibitemOpen
  \bibfield  {author} {\bibinfo {author} {\bibfnamefont {V.}~\bibnamefont
  {Balasubramanian}}, \bibinfo {author} {\bibfnamefont {P.}~\bibnamefont
  {Kraus}}, \ and\ \bibinfo {author} {\bibfnamefont {A.~E.}\ \bibnamefont
  {Lawrence}},\ }\href {\doibase 10.1103/PhysRevD.59.046003} {\bibfield
  {journal} {\bibinfo  {journal} {Phys. Rev.}\ }\textbf {\bibinfo {volume}
  {D59}},\ \bibinfo {pages} {046003} (\bibinfo {year} {1999})},\ \Eprint
  {http://arxiv.org/abs/hep-th/9805171} {arXiv:hep-th/9805171 [hep-th]}
  \BibitemShut {NoStop}%
\bibitem [{\citenamefont {Basu}\ and\ \citenamefont {Wolf}(1982)}]{basu1982}%
  \BibitemOpen
  \bibfield  {author} {\bibinfo {author} {\bibfnamefont {D.}~\bibnamefont
  {Basu}}\ and\ \bibinfo {author} {\bibfnamefont {K.~B.}\ \bibnamefont
  {Wolf}},\ }\href@noop {} {\bibfield  {journal} {\bibinfo  {journal} {Journal
  of Mathematical Physics}\ }\textbf {\bibinfo {volume} {23}},\ \bibinfo
  {pages} {189} (\bibinfo {year} {1982})}\BibitemShut {NoStop}%
\bibitem [{\citenamefont {{Bardeen}}\ and\ \citenamefont
  {{Horowitz}}(1999)}]{bardeen-horowitz1999}%
  \BibitemOpen
  \bibfield  {author} {\bibinfo {author} {\bibfnamefont {J.}~\bibnamefont
  {{Bardeen}}}\ and\ \bibinfo {author} {\bibfnamefont {G.~T.}\ \bibnamefont
  {{Horowitz}}},\ }\href {\doibase 10.1103/PhysRevD.60.104030} {\bibfield
  {journal} {\bibinfo  {journal} {\prd}\ }\textbf {\bibinfo {volume} {60}},\
  \bibinfo {eid} {104030} (\bibinfo {year} {1999})},\ \Eprint
  {http://arxiv.org/abs/hep-th/9905099} {hep-th/9905099} \BibitemShut {NoStop}%
\bibitem [{\citenamefont {Kunze}\ and\ \citenamefont
  {Stein}(1960)}]{Kunze1960}%
  \BibitemOpen
  \bibfield  {author} {\bibinfo {author} {\bibfnamefont {R.~A.}\ \bibnamefont
  {Kunze}}\ and\ \bibinfo {author} {\bibfnamefont {E.~M.}\ \bibnamefont
  {Stein}},\ }\href {http://www.jstor.org/stable/2372876} {\bibfield  {journal}
  {\bibinfo  {journal} {American Journal of Mathematics}\ }\textbf {\bibinfo
  {volume} {82}},\ \bibinfo {pages} {1} (\bibinfo {year} {1960})}\BibitemShut
  {NoStop}%
\bibitem [{\citenamefont {Breitenlohner}\ and\ \citenamefont
  {Freedman}(1982)}]{Breitenlohner:1982jf}%
  \BibitemOpen
  \bibfield  {author} {\bibinfo {author} {\bibfnamefont {P.}~\bibnamefont
  {Breitenlohner}}\ and\ \bibinfo {author} {\bibfnamefont {D.~Z.}\ \bibnamefont
  {Freedman}},\ }\href {\doibase 10.1016/0003-4916(82)90116-6} {\bibfield
  {journal} {\bibinfo  {journal} {Annals Phys.}\ }\textbf {\bibinfo {volume}
  {144}},\ \bibinfo {pages} {249} (\bibinfo {year} {1982})}\BibitemShut
  {NoStop}%
\bibitem [{\citenamefont {Dias}\ \emph {et~al.}(2009)\citenamefont {Dias},
  \citenamefont {Reall},\ and\ \citenamefont {Santos}}]{Dias:2009ex}%
  \BibitemOpen
  \bibfield  {author} {\bibinfo {author} {\bibfnamefont {O.~J.~C.}\
  \bibnamefont {Dias}}, \bibinfo {author} {\bibfnamefont {H.~S.}\ \bibnamefont
  {Reall}}, \ and\ \bibinfo {author} {\bibfnamefont {J.~E.}\ \bibnamefont
  {Santos}},\ }\href {\doibase 10.1088/1126-6708/2009/08/101} {\bibfield
  {journal} {\bibinfo  {journal} {JHEP}\ }\textbf {\bibinfo {volume} {08}},\
  \bibinfo {pages} {101} (\bibinfo {year} {2009})},\ \Eprint
  {http://arxiv.org/abs/0906.2380} {arXiv:0906.2380 [hep-th]} \BibitemShut
  {NoStop}%
\bibitem [{\citenamefont {Amsel}\ \emph {et~al.}(2009)\citenamefont {Amsel},
  \citenamefont {Horowitz}, \citenamefont {Marolf},\ and\ \citenamefont
  {Roberts}}]{Amsel:2009ev}%
  \BibitemOpen
  \bibfield  {author} {\bibinfo {author} {\bibfnamefont {A.~J.}\ \bibnamefont
  {Amsel}}, \bibinfo {author} {\bibfnamefont {G.~T.}\ \bibnamefont {Horowitz}},
  \bibinfo {author} {\bibfnamefont {D.}~\bibnamefont {Marolf}}, \ and\ \bibinfo
  {author} {\bibfnamefont {M.~M.}\ \bibnamefont {Roberts}},\ }\href {\doibase
  10.1088/1126-6708/2009/09/044} {\bibfield  {journal} {\bibinfo  {journal}
  {JHEP}\ }\textbf {\bibinfo {volume} {09}},\ \bibinfo {pages} {044} (\bibinfo
  {year} {2009})},\ \Eprint {http://arxiv.org/abs/0906.2376} {arXiv:0906.2376
  [hep-th]} \BibitemShut {NoStop}%
\bibitem [{\citenamefont {Green}\ \emph {et~al.}(2023)\citenamefont {Green},
  \citenamefont {Hollands}, \citenamefont {Sberna},\ and\ \citenamefont
  {Toomani}}]{greenumeey}%
  \BibitemOpen
  \bibfield  {author} {\bibinfo {author} {\bibfnamefont {S.~R.}\ \bibnamefont
  {Green}}, \bibinfo {author} {\bibfnamefont {S.~}\ \bibnamefont {Hollands}},
  \bibinfo {author} {\bibfnamefont {L.}~\bibnamefont {Sberna}}, 
  \bibinfo {author} {\bibfnamefont {V.}~\bibnamefont {Toomani}},
\ and\ \bibinfo
  {author} {\bibfnamefont {P.}\ \bibnamefont {Zimmerman}},\ }\href {\doibase
  10.1103/PhysRevD.107.064030} {\bibfield  {journal} {\bibinfo  {journal}
  {Phys. Rev. D}\ }\textbf {\bibinfo {volume} {107}},\ \bibinfo {pages} {064030} (\bibinfo
  {year} {2023})},\ \Eprint {http://arxiv.org/abs/2210.15935} \BibitemShut %

\end{document}